\documentclass[aps,nofootinbib,showpacs
]{revtex4}
\usepackage{amsmath}
\usepackage{amsfonts}
\usepackage{amssymb}
\usepackage{graphicx}
\usepackage{epsf}
\newcommand{\ud}{\mathrm{d}}

\begin{document}
\title{Cosmological bootstrap}
\author{V.V.Kiselev} \author{S.A.Timofeev}
\affiliation{
Institute for High Energy Physics,
Moscow region, Protvino, Pobeda 1a, 142281} 
\affiliation{Moscow Institute of Physics and Technology (State University),
Moscow region, Dolgoprudny, Institutsky per.~9, 141701}
\begin{abstract}
A huge value of cosmological constant characteristic for the particle physics
and the inflation of early Universe are inherently related to each other: one
can construct a fine-tuned superpotential, which produces a flat potential of
inflaton with a constant density of energy ${V=\Lambda^4}$ after taking into
account for leading effects due to the supergravity, so that an introduction
of small quantum loop-corrections to parameters of this superpotential
naturally results in the dynamical instability relaxing the primary
cosmological constant by means of inflationary regime. The model
phenomenologically agrees with observational data on the large scale
structure of Universe at $\Lambda\sim 10^{16}$~GeV.
\end{abstract}
\maketitle

\section{Introduction}

In modern cosmology there are actual problems of cosmological constant
\cite{Weinberg-RMP} and inflation
\cite{i-Guth,i-Linde,i-Albrecht+Steinhardt,i-Linde2,inflation}. Namely, if we
consider the first among the above mentioned issues, then in the framework of
quantum field theory we could naturally expect that a scale of vacuum energy
is determined by a value characteristic for interactions of elementary
particles, hence, the cosmological constant would get the huge value
comparable with the Planck mass or, at least, with a scale of gauge symmetry
breaking in the particle physics such as a grand unification and the
electroweak symmetry or, finally, with the characteristic values of
quark-gluon condensates. However, in all of listened cases the cosmological
constant would essentially exceed a constraint extracted from observations of
the anisotropy of cosmic microwave background radiation (CMBR)
\cite{WMAP5-1,WMAP5-2,WMAP}, the inhomogeneity of matter distribution in the
Universe (LSS -- large scale structure) \cite{BAO}, the dependence of
brightness for the type Ia supernovas on the red shift (SnIa -- supernova Ia)
\cite{Riess:2004nr,Riess:2006fw,Astier:2005qq,WoodVasey:2007jb}, because, in
accordance with these data, the admissible value of vacuum energy density has
got the scale of $10^{-3}$~eV, that contradicts to the concept of
characteristic energies corresponding to the field interactions. Therefore,
in the literature, there are various discussions on mechanisms transforming
the initial huge-valued cosmological constant to the reduced value close to
that of observed.

Among such the models we emphasize the renormalization group approach with a
``running'' cosmological constant evolving versus the rate of Universe
expansion, i.e. versus the Hubble parameter
\cite{ShapiroSola,ShapiroSola-2,Guberina,Shapiro:2004,Bilic,Sola,
rev-CC,Shapiro:2009dh,Bauer:2009ke}. In this way, a minimum of effective
potential, i.e. the density of vacuum energy, being invariant under the
renormalization group transformations, depends on the coupling constants
given by values of fields related with the Hubble rate, that leads to a slow
logarithmic evolution of cosmological constant in agreement with the
equations of renormalization group.

In another approach, among all of non-renormalizable theories including the
Einstein gravitation, one specifies a kind of asymptotically safe theories
\cite{Weinberg-safe} with the following properties: for a countable number of
local operators of non-renormalizable theory with arbitrary coefficients one
isolates dimensional factors in the form of degrees of a scale, so that
overall dimensionless constants or ``charges'' satisfy equations of a
renormalization group with \textit{a final number of fixed points},
therefore, the theory \textit{asymptotically} gets a predictive power in
vicinity of attractive fixed points. The asymptotically safe gravitation can
include the evolution of cosmological term, say, to its zero value as well as
the inflation \cite{Weinberg:2009wa}.

In a new approach by J.D.~Barrow and D.J.Show \cite{BS1,BS2,BS3}, the
cosmological constant is considered as a field, so that its value is
extracted from the principle of extremal action restricted by causality. As a
general consequence, one finds that the scale of cosmological constant is
naturally determined by the inverse age of present Universe, in Planckian
units. In the case of general relativity describing the evolution of
homogeneous isotropic Universe, there is a falsifiable connection of
cosmological constant to the spatial curvature satisfying the present limits
observed. The curvature is intrinsically related with an amount of inflation,
so that a distribution of probability for the inflation consistent with the
properties of our Universe has got a narrow peak in vicinity of cosmological
constant fixed by an appropriate small scale.

In \cite{K1,K2,K3} V.~Emelyanov and F.R.~Klinkhamer considered a mechanism
compensating  a primary cosmological constant due to specific vector fields,
so that the Minkowsky spacetime with zero value of cosmological constant can
be an attractor of dynamical equations. In other approach, the cosmological
constant relaxes by gradually passing different regions of its potential,
i.e. by moving from a plateau to plateau \cite{KV}.

One also has considered a possibility of application for the quantum mechanical
mechanism of tiny mixing of two non-stationary states (the so called seesaw)
with different densities of energy as given by the particle physics. Then,
the mixing results in the stationary vacuum getting the cosmological constant
of reduced scale \cite{Chalmers,Grav-seesaw,KT2,KT1,KT-pp,
Enqvist:2007tb,Banks-I,Banks-Heretics}.

Next, the inflation model, solving a lot of problems for the observational
cosmology, leads to a potential of scalar inflaton, that is characterized by
the mass $m\approx 1.5\times 10^{13}$ GeV, a vacuum expectation value
essentially exceeding the Planck scale of energy, a tiny value of
self-interaction constant (the constant of quartic self-action for the field,
$\lambda\sim 10^{-13}$) and a flat plateau at the magnitude of energy density
$\Lambda^4$ at $\Lambda\sim 10^{16}$ GeV. In this respect, one has got the
question on the naturalness of such the exotic potential (see exhaustive
review on the inflation relation to the particle physics and a mechanism of
reheating and thermalization of Universe after the
inflation\footnote{Realistic models of low-energy inflation taking into
account for constraints coming from the primordial nucleosynthesis, the
anisotropy of CMBR and the inhomogeneity of matter distribution in the LSS of
Universe are presented in
\cite{Allahverdi:2006iq,Allahverdi:2006cx,Allahverdi:2006we}, wherein the
supersymmetric version of Standard Model is explored with the usage of flat
directions in the superpotential.} in \cite{Mazumdar:2010sa}).

To solve the problem of inflaton naturalness, at present some studies are
actively focused to a model of Higgs scalar $\Phi_H$ in the Standard model
with a non-minimal coupling to the gravity (to the scalar curvature $R$) in
the form of lagrangian $L_\mathrm{int}=\xi R\,\Phi_H^\dag \Phi_H$, where the
coupling constant has got a value about $\xi\sim 10^4$. Further, under a
conformal transformation to an effective field of inflaton minimally coupled
to the gravity, one gets rather a flat potential with the necessary magnitude
of plateau \cite{Barvinsky,Bezrukov,DeSimone,Barvinsky2,Bezrukov2,Burgess,
Barbon,CervantesD,CervantesD-GUT,Barvinsky3}. In this way, an account for
corrections calculated within the method of appropriate renormalization
group, leads to strict constraints on the mass of Higgs boson: $135.6 \mbox{
GeV} < m_H < 184.5 \mbox{ GeV}$.

The Higgs boson minimally coupled to the gravity ($\xi\to 0$) could play an
essential role in the dynamics of early Universe \cite{KT-dec}. Indeed, there
is a critical value of Higgs boson-mass, which is approximately equal to
$m_H^\mathrm{crit.}\approx 153\pm 3$ GeV after an account for two-loop
corrections \cite{KT-RG}, so that at supercritical values of the mass, the
Higgs scalar is not able to cause the inflation of Universe. If the Higgs
particle is the only scalar field in the theory up to Planckian scales of
energy, then the Higgs boson of subcritical mass is forbidden, since the
inflation caused by such the field would generate the Big Bang Universe with
a large scale structure of matter exceptionally different from the observed
one. At supercritical values of Higgs boson-mass, a distribution of matter
inhomogeneity would be determined by finely tuned initial data (that would be
avoided due to the inflation). Thus, the subcritical values of Higgs
boson-mass will inevitably require the introduction of scalar field of
inflaton additional to the Standard model, and the inflaton should
dynamically provide us with the formation of necessary properties of large
scale structure in our Universe \cite{KT-PoS-Paris}.

Cosmological constraints on the mass of Higgs boson can be obtained in other
approaches like a consideration of the scalar in companion with another field
as was done in \cite{Perv1}, wherein the constraint in the form of $m_H<134$
GeV has been derived.

In \cite{KYY} a model of inflation was constructed in the framework of
supergravity by means of setting an appropriate kind of Kahler potential with
an additional symmetry keeping the Kahler potential to be independent of an
imaginary part of scalar field. However, on a more deep level of superstring
theory one cannot construct a way leading to models analogous to that of
\cite{KYY}. In \cite{KKLT} one has offered a realistic model of inflation in
the framework of superstrings, so that problems of instability caused by the
compactification of extra dimensions can be removed, and a vacuum is shifted
from an initial anti-de Sitter state to a minimum of inflaton potential with
a positive energy, i.e. to the de Sitter vacuum. Nevertheless, to our
opinion, the development of inflation can occur at such densities of energy,
whereat the supergravity is certainly broken, and a consistent field theory
at the sub-Planckian energies should be based on the principle of
renormalizability, while potentials derived in the framework of complete
exact supergravity include terms of higher dimensions in the Newtonian
constant $G$, which are not renormalizable. Therefore, such the terms
presumably should be canceled after the breaking the supergravity, so that in
the effective potential one has to hold the terms consistent with the
requirement of renormalizability only, i.e. the terms being under control
with respect to quantum loop-corrections. Otherwise, the field theory at
energies below the Planck scale loses any predictable power. In this respect,
one should to restrict himself by considering the supergravity corrections to
the potential within limits of leading terms linear in the Newtonian constant
$G$ as was done by S.~Weinberg in \cite{W3}, combining this limit with the
constraint on the degree of field self-action, that should be not greater
than 4.

One can consider the problems of cosmological constant and inflationary
dynamics beyond the general relativity, for instance, in the framework of
conformal general relativity \cite{Perv2}.

In the present paper we suggest to look at the problem of relaxing the
cosmological constant and the naturalness of parameters in the potential of
scalar field generating the inflation, from the unified point of view based
on a phenomenological introduction of dynamical field with the flat potential
$V=const.$ (up to contributions restricted by a power of inverse Planck
mass), so that this potential value gives the primary cosmological constant.
Then, the primary effective values of both the mass and self-action constant
for the dynamical field are equal to zero. Such the fields can naturally
appear in superstring theory and they are called modules: such the fields
determine ``flat directions'' of effective potential. Then, we study whether
the flat potential is stable under perturbations caused by quantum
loop-corrections given by subleading terms in the inverse Planck mass.

So, we introduce such the field as a scalar component of chiral superfield in
the supersymmetric theory. Further, we investigate the problem to determine
nontrivial parameters of superpotential for the field, so that after taking
into account for leading supergravity corrections in weak gravitational
fields, i.e. \textit{corrections linear in the gravitational constant}, the
resulting potential would be flat, hence, the field would actually be the
module. Such the procedure of determining the parameters of primary
superpotential we call the cosmological bootstrap, because it fixes the
initial mass and self-action constant of field in accordance with the
requirement of canceling those terms by leading contributions due to the
supergravity.

In this way, we assume that the scale of potential plateau is essentially
less than the Planck mass, but it is large enough to agree with the concept
of particle interactions: we put the scale to be close to an energy of grand
unification of gauge interactions. Then, we find that the bootstrap is
possible, i.e. the fine tuning can be specified, and the initial parameters
of field are close to the mass and self-action constant for the inflaton,
generating the observed large scale structure of Universe.

Next step of study is a small perturbations of fine tuning that leads to
breaking of bootstrap for the mass and self-action of field as caused by
higher corrections in the inverse Planck mass appearing due to loops taking
into account for propagation of heavy particles with masses about the Planck
mass. These quantum loop corrections result in the dynamical instability of
primary cosmological constant, i.e. in the instability of initial flat
potential, with the further relaxation of cosmological constant during the
inflation generated by the field primary marked as the module. Then, the
inflation parameters are naturally given by the breaking of bootstrap, and
they agree with the observed values by the order of magnitude. It is
important to emphasize that the mass, vacuum expectation value of the field
as well as its self-action constant are actually determined by the
introduction of single dimensional parameter being the scale of primary
cosmological constant, which is natural for the particle physics.

\section{Bootstrap}

In the theory of gravity with the dimensional coupling constant $G$ it would
be natural to expect that the vacuum energy $\rho_G$ giving the cosmological
constant, is determined by the Planck scale\footnote{We use the reduced
Planck mass.} ${\tilde m}_\mathrm{Pl}=1/\sqrt{8\pi G}\approx
2.4\times10^{18}$ GeV, so that
\begin{equation}\label{grav}
    \rho_G=M^4,
\end{equation}
where $M\sim {\tilde m}_\mathrm{Pl}$. However, the supersymmetry can be
dynamically related to another scale of energy $\Lambda$, which will be set
much less than the Planck mass,
\begin{equation}\label{hier}
    \Lambda\ll M.
\end{equation}
In a phenomenological point of view, i.e. without a presentation of any
mechanism for the introduction of scale $\Lambda$ in addition to ${\tilde
m}_\mathrm{Pl}$, we will assume that in the model there are two dimensional
quantities with the definite hierarchy of (\ref{hier}). In this way we will
naturally suggest that the local supersymmetry, i.e. the supergravity leads
to the vacuum energy equal to
\begin{equation}\label{super}
    \rho_S=\Lambda^4.
\end{equation}
Such the finite renormalization\footnote{In this paper we do not consider
questions on both a regularization of infinities and a renormalization
group.} from (\ref{grav}) to (\ref{super}) is really possible in the
supergravity, since the introduction of superpotential in the form of
constant
\begin{equation}\label{W0}
    W_0= i \omega_0^3,
\end{equation}
leads to the additional term\footnote{Following S.~Weinberg \cite{W3}, we restrict the
consideration by corrections linear in the gravitational constant $G$.} in
the density of vacuum energy, so that (see \cite{W3})
\begin{equation}\label{cb-1}
    V_0=\rho_G-24\pi G|W_0|^2=M^4-24\pi G\omega_0^6,
\end{equation}
while the condition of $V_0=\rho_S$ is the definition for the scale
$\omega_0$. The above procedure presents the ``nil'' step of bootstrap being
the consequent determination of theory parameters in terms of primary
quantities, particularly, the determination of constant term in the
superpotential in terms of two primary scales\footnote{Here we certainly
suggest that all of quantities with the same order of magnitude are
equivalent, of course.} $M$ and $\Lambda$. In this way, our goal is the
description of primary cosmological constant as the flat potential.

Further, in the framework of supersymmetry the vacuum energy is determined by
the potential
\begin{equation}\label{spoten}
    V_S=\left|\frac{\partial W}{\partial\Phi}\right|^2,
\end{equation}
where, for the case of cosmological constant, i.e. for the potential independent
of the field $\Phi$, the superpotential should be written in the form
\begin{equation}\label{cb-2}
    W\mapsto W_1=i\omega_0^3+\Lambda^2\Phi,
\end{equation}
so that
\begin{equation}\label{spot2}
    V_S=\Lambda^4.
\end{equation}
The account for the supergravity with the linear terms in the Newtonian
constant $G$ leads to the potential of general form (see  \cite{W3})
\begin{equation}\label{pot}
    V=\rho_G+\left|\frac{\partial W}{\partial\Phi}\right|^2-
    24\pi G \left|W-\frac{1}{3}\Phi\frac{\partial
    W}{\partial\Phi}\right|^2+\frac{16\pi G}{3}|\Phi|^2
    \left|\frac{\partial W}{\partial\Phi}\right|^2.
\end{equation}
Then the substitution of superpotential in the form of (\ref{cb-2}) into eq.
(\ref{pot}) gives
\begin{equation}\label{pot1}
    V\mapsto V_1=(M^4-24\pi
    G\omega_0^6)+\Lambda^4-\frac{1}{2}\,\phi^2\,\frac{16\pi
    G}{3}\Lambda^4,
\end{equation}
if one puts the field $\phi=\Phi\sqrt{2}$ to be real (a consideration of
general case is presented in Appendix \ref{appen}). Moreover, for the sake of
simplicity in what follows, we suggest that the potential accounting for the
supergravity corrections of (\ref{pot}), possesses the symmetry with respect
to discrete operation of reflection $\phi\leftrightarrow -\phi$, i.e. it does
not include odd degrees of field $\phi$. At this step of bootstrap, the
cosmological constant is given by the energy density $V_S$ in (\ref{spot2}),
if we slightly adjust the parameter $\omega_0$, so thats
\begin{equation}\label{cb-omega}
    M^4-24\pi G\omega_0^6=0.
\end{equation}
However, in addition to the constant density of energy, the potential of
(\ref{pot1}) includes the term depending on the field. Moreover, the
supergravity corrections leads to the instability of cosmological term. The
instability actually signalizes that the supergravity renormalizes the term
quadratic in the field, hence, the interaction quadratic in the field should
be introduced at the stage of primary superpotential $W$. In addition, we see
that in the theory with the global supersymmetry the module loses its main
property in the theory with the local supersymmetry, i.e. in the
supergravity. Therefore, we challenge the problem to determine a nontrivial
superpotential, which describes the module after the account for the
supergravity corrections in the form of (\ref{pot}).

Following such the bootstrap approach, we write down the superpotential with real parameters in the form
\begin{equation}\label{spot-x}
    W=i\omega_0^3+\Lambda^2\Phi+\frac{i}{2}\,\mu_0\Phi^2+\frac{g_0}{3}\,\Phi^3.
\end{equation}
Then, \textit{with the accuracy up to terms quartic in the field}, i.e.
accounting for the renormalizable terms, we get the potential $V=\hat
V\Lambda^4$ with
\begin{equation}\label{pot-x}
    \hat V=1-\frac{1}{2}\,\hat\mu^2\hat \phi^2+
    \frac{1}{4}\,\hat\lambda\hat\phi^4,
\end{equation}
wherein we take into account the cancelation due to (\ref{cb-omega}) at the
``nil'' step of bootstrap, and the following notations for the dimensionless
quantities with hats are adapted:
\begin{equation}\label{hats}
    \begin{array}{lcl}\displaystyle
      \hat \phi^2=\frac{32\pi G}{3}\,|\Phi|^2, & &
      \displaystyle
      \hat g_0=-\frac{3}{16\pi G\Lambda^2}\,g_0, \\[3mm] \displaystyle
      \hat\omega_0^6=\frac{16\pi G}{3\Lambda^4}\,\omega_0^6, & &
      \displaystyle
      \hat\mu_0^2=\frac{3}{16\pi G\Lambda^4}\,\mu_0^2,
    \end{array}
\end{equation}
so that
\begin{equation}\label{param}
    \begin{array}{l}\displaystyle
      \hat\mu^2=1-\hat\mu_0^2+2\hat g_0+
    \frac{3}{2}\,\hat \omega_0^3\hat\mu_0, \\[3mm]
    \displaystyle
      \hat\lambda=\hat g_0^2-2\hat
    g_0+\frac{7}{8}\,\hat\mu_0^2.
    \end{array}
\end{equation}

Potential (\ref{pot-x}) represents the constant density of energy, if the
bootstrap relations are in action,
\begin{equation}\label{cb-x}
    \hat\mu^2=\hat\lambda=0.
\end{equation}

Accounting for $\hat \omega_0^6\gg1$, we can easily find solutions of
equations (\ref{cb-x}) for parameters $\hat \mu_0$ and $\hat g_0$: there are
two sets corresponding to solutions of quadratic equation for $\hat g_0$,
\begin{equation}\label{hat-g}
    \hat g_0=1 \pm \sqrt{1 - \frac{7}{8}\,\hat\mu_0^2}.
\end{equation}
Namely,
\begin{equation}\label{sets}
    \begin{array}{ll}\displaystyle
      \hat\mu_0^{I}\approx -\frac{2}{3}\frac{1}{\hat\omega_0^3}, &\displaystyle
      \qquad \hat\mu_0^{II}\approx -\frac{10}{3}\frac{1}{\hat\omega_0^3},
      \\[3mm] \displaystyle
      \hat g_0^{I}\approx \frac{7}{16}\left\{\hat\mu_0^{I}\right\}^2, & \displaystyle
      \qquad \hat g_0^{II}\approx
      2-\frac{7}{16}\left\{\hat\mu_0^{II}\right\}^2.
    \end{array}
\end{equation}
Sub-leading corrections in (\ref{sets}) can be presented in the form of
expansion in \textit{even} degrees of ratio $\Lambda/{\tilde m}_\mathrm{Pl}$.
Since the coefficients of such the expansion are strictly definite, we can
see that, first, the ``fine tuning'' of bootstrap parameters is required and,
second, the mistuning leads to breaking of bootstrap. The mechanism of
breaking we study in the present paper, is the following: the introduction of
corrections in powers of $\Lambda/{\tilde m}_\mathrm{Pl}$ to the ``bare''
superpotential causes the dynamical instability of primary cosmological
constant, that leads to the inflationary expansion of Universe.

\begin{figure}
  \includegraphics[width=10cm]{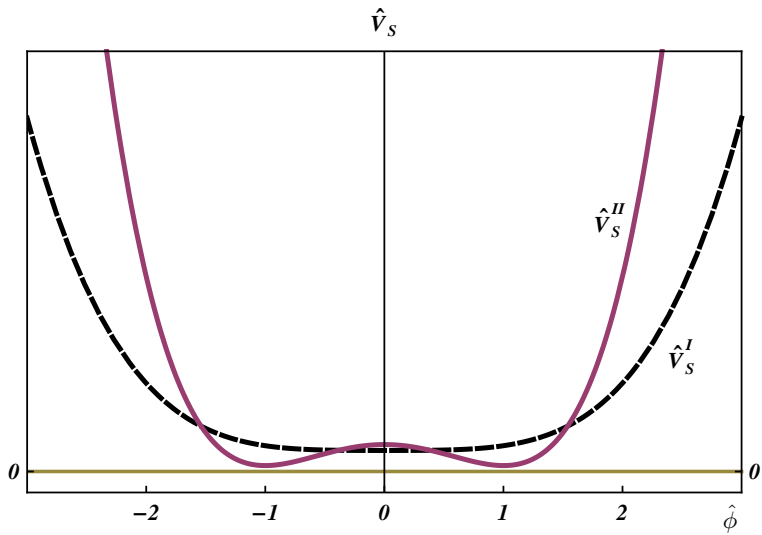}\\[-5mm]
  \setlength{\unitlength}{1mm}
  \begin{picture}(70,5)
  \put(80,0){$\hat\phi$}
  \end{picture}
  \caption{The primary potential of scalar field for the sets I and II
   (in arbitrary units).}\label{fig-s}
\end{figure}

Sets (\ref{sets}) correspond to different hierarchies for the values of
initial constant in the field self-action, $\hat g_0$, since $\hat
g_0^{II}\gg \hat g_0^{I}$, but they also lead to essentially different
initial potentials of field in the framework of supersymmetry according to
(\ref{spoten}). Indeed, the primary potential $V_S=\hat V_S\,\Lambda^4$ can
be written in the form
\begin{equation}\label{spoten-2}
    \hat
    V_S=1+\frac{\hat\mu_S^2}{2}\,\hat\phi^2+\frac{\hat\lambda_S}{4}\,\hat\phi^4,
\end{equation}
where
\begin{equation}\label{mu-s}
    \begin{array}{l}
      \hat\mu_S^2=\hat\mu_0^2-2\hat g_0, \\
      \hat\lambda_S =\hat g_0^2.
    \end{array}
\end{equation}
For (\ref{sets}), the initial potential is stable: $\hat g_0^2>0$. Moreover,
taking into account for the approximation $\hat \omega_0^6\gg1$, we can write
down the explicit form
\begin{equation}\label{sponten-3}
    \begin{array}{l}\displaystyle
      \hat V_S^{I}\approx 1+\left\{\frac{1}{4}\,\hat\mu_0^{I}\right\}^2\,\hat\phi^2+
      \left\{\frac{7}{32}\,\hat\mu_0^{I}\right\}^2\hat\phi^4,  \\[4mm]
      \displaystyle
      \hat V_S^{II}\approx
      (1-\hat\phi^2)^2+\frac{1}{2}\,\left\{\hat\mu_0^{II}\right\}^2.
    \end{array}
\end{equation}
The character of dependencies of initial potentials for the bootstrap sets I
and II is shown in fig. \ref{fig-s}, wherefrom we see that the set I
corresponds to the situation, when the breaking of supersymmetry corresponds
to the vacuum energy density $\Lambda^4$, and the vacuum expectation value of
field is equal to zero, while the set II involves the reduced value of field
contribution to the supersymmetry breaking, that is accompanied by the
spontaneous breaking of symmetry with respect to the discrete inversion
$\phi\leftrightarrow-\phi$ suggested above. In addition, the set II is
characterized by ``natural'' scaling for the values of both the mass and the
constant of self-action, while for the set I these parameters take reduced
values.

The breaking of bootstrap relations in (\ref{cb-x}) due to the quantum
loop-corrections leads to instability of primary cosmological constant $\hat
V=1$ as well as to the inflation of Universe, if the minimum of potential
corresponds to the scaled density of energy negligibly less than the unit. In
this case, the primary cosmological constant relaxes from the huge value to
that of we will put equal to zero\footnote{Otherwise, one should introduce
the new dynamical field, which again causes the instability of remnant
nonzero density of energy, if it is positive. A negative value of remnant
density of energy would lead to the collapse of Universe, that is not
observed. See the discussion of this issue in the next section.}.

\section{Phenomenological analysis}

The breaking of bootstrap relations is caused by the variation of parameters
$\omega_0$, $\mu_0$ and $g_0$. In this section we do not specify any
mechanism of breaking, but we will generally accept that the quantum
corrections are expandable in the small ratio of $\Lambda/{\tilde
m}_\mathrm{Pl}$. In this way, we find some natural constraints.

First, corrections to $\omega_0$ contribute to the vacuum energy, so that to
the leading approximation it gets the contribution of the order of $\delta
V_0\sim\Lambda^4$, hence, $\delta\hat\omega_0^6\sim \mathcal{O}(1)$, in
accordance to the definition of $\delta V_0\sim
\Lambda^4\delta\hat\omega_0^6$. By construction,
\begin{equation}\label{delta-omega}
    \hat\omega_0^3\sim
    \left(\frac{{\tilde m}_\mathrm{Pl}}{\Lambda}\right)^2
    \quad\Rightarrow\quad
    \frac{\delta\hat\omega_0^3}{\hat\omega_0^3}\sim
    \left(\frac{\Lambda}{{\tilde m}_\mathrm{Pl}}\right)^4.
\end{equation}
The same correction can lead to the variation of parameter
$\hat\mu_0\sim\hat\omega_0^{-3}$, so that
\begin{equation}\label{delta-mu1}
    \frac{\delta'\hat\mu_0}{\hat\mu_0}\sim
    \left(\frac{\Lambda}{{\tilde m}_\mathrm{Pl}}\right)^4,
\end{equation}
and the consequent variation of self-action constant $\delta'\hat g_0$ will
appear, too, but such the transition of variations with the valid relations
of bootstrap will not generally lead to the breaking of bootstrap itself, and
we suggest that the parameters have got independent sources of corrections.

Second, the parameters of bootstrap have got the following orders of
magnitude:
\begin{equation}\label{mu-g}
    \hat\mu_0\sim \left(\frac{\Lambda}{{\tilde m}_\mathrm{Pl}}\right)^2,\qquad
    \hat g_0\sim
    \mathcal{O}(1)+\left(\frac{\Lambda}{{\tilde m}_\mathrm{Pl}}\right)^4,
\end{equation}
hence, the leading corrections can be written in the form
\begin{equation}\label{delta-mu-g}
    \delta\hat\mu_0\sim
    \left(\frac{\Lambda}{{\tilde m}_\mathrm{Pl}}\right)^{2+q},\qquad
    \delta\hat g_0\sim
    \left(\frac{\Lambda}{{\tilde m}_\mathrm{Pl}}\right)^{4+\tilde q},
\end{equation}
where the integer degrees are $q,\tilde q\geqslant 0$. The bootstrap breaking
according to (\ref{param}) is caused by nonzero values of quantities
\begin{equation}\label{break}
    \begin{array}{l}\displaystyle
      \hat\mu^2=-2\hat\mu_0\delta\hat\mu_0+2\delta\hat g_0+\frac{3}{2}\,\hat\omega_0^3\delta\hat\mu_0,
      \\[3mm] \displaystyle
      \hat\lambda=2(\hat g_0-1)\delta\hat
      g_0+\frac{7}{4}\,\hat\mu_0\delta\hat\mu_0,
    \end{array}
\end{equation}
so that
\begin{equation}\label{break-2}
    \begin{array}{clcl}
      \hat\mu^2&\displaystyle
      \sim \left(\frac{\Lambda}{{\tilde m}_\mathrm{Pl}}\right)^q &+&
      \displaystyle
      \mathcal{O}(1)\left(\frac{\Lambda}{{\tilde m}_\mathrm{Pl}}
      \right)^{4+\tilde q},
      \\[5mm]
      \hat\lambda &\displaystyle
      \sim \left(\frac{\Lambda}{{\tilde m}_\mathrm{Pl}}\right)^{4+\tilde q}
      &+& \displaystyle
      \mathcal{O}(1)\left(\frac{\Lambda}{{\tilde m}_\mathrm{Pl}}
      \right)^{4+q}.
    \end{array}
\end{equation}
Then, the minimum of potential $\hat V$ at $\hat\phi^2=\hat\mu^2/\hat\lambda$
gets the value
\begin{equation}\label{Vmin}
    \hat V_\mathrm{min}=1-\frac{\hat\mu^4}{4\hat\lambda},
\end{equation}
and the relaxation of primary cosmological constant meaning the cancelation
of contributions in the density of vacuum energy of the order of $\Lambda^4$,
is possible, if only
\begin{equation}\label{relax}
    \frac{\hat\mu^4}{4\hat\lambda}\sim\mathcal{O}(1).
\end{equation}
The condition of (\ref{relax}) implies that, in the case of $q\geqslant\tilde
q$,
\begin{equation}\label{rel1}
    2q=4+\tilde q,
\end{equation}
so that there is the finite set of values for the correction degrees
\begin{equation}\label{rel1-1}
    \{q,\tilde q\}\mapsto \{2,0\},\;\{3,2\},\;\{4,4\}.
\end{equation}
If $q<\tilde q$, then there is the solution $q=4$ with an arbitrary value of
$\tilde q\geqslant 5$. In all of those cases, the scaling behavior of
potential parameters is reduced to
\begin{equation}\label{break-3}
    \hat \mu^2\sim \left(\frac{\Lambda}{{\tilde m}_\mathrm{Pl}}\right)^q,\qquad
    \hat \lambda\sim \left(\frac{\Lambda}{{\tilde m}_\mathrm{Pl}}\right)^{2q},
\end{equation}
where $q=\{2,3,4\}$.

Thus, we get the general phenomenological description of corrections
responsible for the breaking of cosmological bootstrap. Remember, the
corrections to the parameter $\hat \omega_0^6$ are reduced to the variation
of $\delta\hat V_\mathrm{min}\sim\mathcal{O}(1)$, so that one gets the
complete cancelation of terms in the vacuum density of energy of the order of
$\Lambda^4$, i.e. up to the required accuracy the condition $\hat
V_\mathrm{min}=0$ would be valid. This constraint is natural for the
bootstrap construction, because the ``survive'' of terms like $\Lambda^4$ in
the vacuum density of energy would lead to the necessity of additional
introduction of secondary field of module, hence, if we consider the ultimate
physical field of module, then the cancelation of cosmological term in the
bootstrap breaking is simply the definition of such the field.

Finally, we adapt that the breaking of cosmological bootstrap for the field
of module results in the scaling potential
\begin{equation}\label{fin-pot}
    \hat V=\left(1-\frac{\hat\mu^2}{4}\,\hat\phi^2\right)^2.
\end{equation}
Therefore, for the physical real field $\phi=\sqrt{2}|\Phi|$ the mass and
constant of self-action have the forms
\begin{equation}\label{fin-mass}
    \begin{array}{l}\displaystyle
      m^2=\frac{32\pi G}{3}\,\Lambda^4\hat\mu^2, \\[4mm] \displaystyle
      \lambda =\left(\frac{8\pi G}{3}\,\Lambda^2\right)^2\hat\mu^4,
    \end{array}
\end{equation}
while, by the order of magnitude,
\begin{equation}\label{fin-mass2}
    \begin{array}{l}\displaystyle
      m\sim {\tilde m}_\mathrm{Pl}\,\left(\frac{\Lambda}{{\tilde m}_\mathrm{Pl}}\right)^{2+q/2}, \\[4mm]
      \displaystyle
      \lambda \sim \left(\frac{\Lambda}{{\tilde m}_\mathrm{Pl}}\right)^{4+2q},
    \end{array}
\end{equation}
and the vacuum expectation value is equal to
\begin{equation}\label{fin-v}
    v=\frac{m}{\sqrt{2\lambda}}\sim
    {\tilde m}_\mathrm{Pl}\,\,\left(\frac{{\tilde m}_\mathrm{Pl}}{\Lambda}\right)^{q/2}.
\end{equation}

The inflaton mass is strictly constrained by the observational
data\footnote{We follow the analysis performed by us in the method of
driftage of attractor in the phase plain, that describes the inflationary
dynamics \cite{KT3,Mexicans,KT4}.}, so that, putting $m/{\tilde
m}_\mathrm{Pl}\sim 10^{-5}$, we get the following characteristic values of
model parameters by the order of magnitude:

\begin{center}
\begin{tabular}{p{20pt}p{55pt}p{55pt}p{40pt}}
  $q$ & \centering $\displaystyle\frac{\Lambda}{{\tilde m}_\mathrm{Pl}}$~~ & \centering
  $\lambda$~~ & $\displaystyle\frac{v}{{\tilde m}_\mathrm{Pl}}$
  \\[3mm]
  \hline
  2 & $2\times 10^{-2}$ & $4 \times 10^{-14}$ & ~$50$ \\
  3 & $4 \times 10^{-2}$ & $5 \times 10^{-14}$ & $150$ \\
  4 & $6 \times 10^{-2}$ & \centering$ 10^{-15}$ &   $300$ \\
\end{tabular}
\end{center}

The analysis of WMAP data on the anisotropy of cosmological microwave
background radiation by 5 years of operation \cite{WMAP5-1} in the framework
of inflaton with the Higgsian kind of potential \cite{KT3} leads to the
following constraints on the vacuum expectation value of the field:
\begin{itemize}
  \item in the new scenario of inflation (the field slowly rolling down
      to the minimum from the state close to the instable local maximum
      at $\phi=0$, the ``hilltop'' inflation)
\begin{equation}\label{fin-v2}
    \frac{v}{{\tilde m}_\mathrm{Pl}}\geqslant 10 ,
\end{equation}
  \item in the scenario of chaotic inflation (the evolution of field to
      the minimum from $|\phi|>v$)
\begin{equation}\label{fin-v3}
    \frac{v}{{\tilde m}_\mathrm{Pl}} \geqslant 100,
\end{equation}
\end{itemize}
so that among the admissible values of parameter characterizing the
contribution of corrections we have to prefer for $q=2$, since the large
vacuum expectation values correspond to the potential, which degenerates to
the limit of $V\sim \phi^2$, that actually lies at the edge of region with
the  1$\sigma$-confidence level. At $q=2$, the scale of primary cosmological
constant is $\Lambda\sim 5\times 10^{16}$ GeV, that, in fact, is consistent
with the suggestion on the scale, corresponding to the grand unification of
gauge interactions.

\begin{figure}[ht]\setlength{\unitlength}{0.75mm}
\begin{picture}(110,80)
\put(0,0){  \includegraphics[width=110\unitlength]{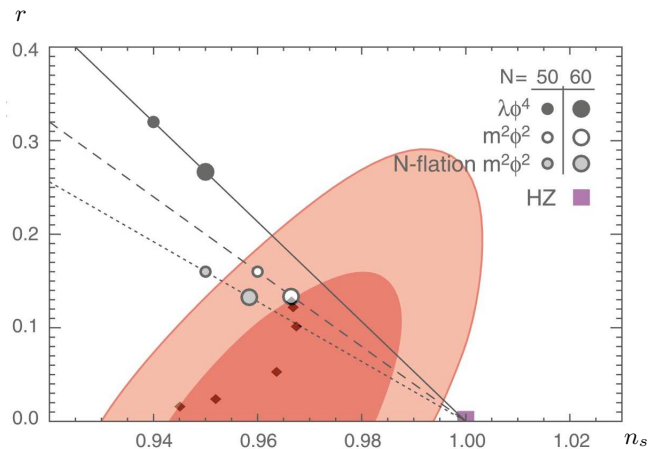}}
\put(111,2){$n_s$}
\put(3,77){$r$}
\end{picture}
  \caption{The comparison of data observed with the 1$\sigma$ and 2$\sigma$ accuracy
  for the spectral index of scalar fluctuations of energy density $n_s$ and
  relative contribution of tensor fluctuations $r$ from \cite{WMAP} with
  predictions of inflation models for the quadratic and quartic self-action,
  and the model with $N$ scalar fields (N-flation), as well as with the limit of
  scale invariant fluctuations by Harrison--Zeldovich (HZ) at the horizon,
  shifted from the end of inflation by e-foldings of scale factor, $N=50$ and $N=60$.
  The predictions of new scenario of inflation are shown by rhombuses at $N=60$ and
  vacuum expectation values of inflaton $ v/{m}_\mathrm{Pl}=2.5,\, 2.8,\,4,\, 7.4,\,
  24.6,\,54$ (in course of growing $r$).
  }\label{fig-ns-r}
\end{figure}

This point gets a more significant confirmation after the account of WMAP
data during the 7-years operation  \cite{WMAP}, which exclude the chaotic
scenario of inflation (with $|\phi|>v$) at the 1$\sigma$-confidence level and
give the followings:
\begin{itemize}
  \item the mass of inflaton equals
  \begin{equation}\label{inf-mass}
    m_\mathrm{inf}\approx (1.30-1.74)\times 10^{13}\mbox{ GeV,}
  \end{equation}
  \item the vacuum expectation value $\langle \phi\rangle =v$ is
      constrained by
  \begin{equation}\label{inf-vev}
    2.5\, m_\mathrm{Pl} < v < 54\, m_\mathrm{Pl},
  \end{equation}
  where $m_\mathrm{Pl}=\sqrt{8\pi}\,{\tilde m}_\mathrm{Pl}$ is the Planck
  mass,
  \item the ``hilltop'' inflation is actual, only, the fields rolls from
      the plateau near the ``nil'' value of field to the potential
      minimum, while the plateau magnitude is given by the parameter
      $V(0)=V_\mathrm{hill}=\Lambda_\mathrm{hill}^4$,
  \begin{equation}\label{hill}
    \Lambda_\mathrm{hill}= (1.2-6.0)\times 10^{16}\mbox{ GeV.}
  \end{equation}
\end{itemize}
The analysis is shown in fig. \ref{fig-ns-r}, wherefrom we see that the
preferable value is $q=2$.

Thus, we expect that in the viable model of corrections, the parameters of
primary superpotential are shifted in accordance with the rules
\begin{equation}\label{shifts-prim}
    \delta\mu_0^2\sim-\mu_0^2
    \left(\frac{\Lambda}{{\tilde m}_\mathrm{Pl}}\right)^2,\qquad
    \delta g_0 \sim \pm\left(\frac{\Lambda}{{\tilde m}_\mathrm{Pl}}\right)^6,
\end{equation}
where the sign of $\delta g_0 $ corresponds to the stability of potential
($\tilde\lambda>0$) for the sets I and II, respectively.

\section{1-loop structure}
In the simplest case, after the supersymmetry breaking the model includes the
single light scalar real field, whereas the notion ``light'' implies that the
field mass is essentially less that the scale of primary cosmological
constant:
$$m\ll \Lambda.$$
Moreover, the light fields are the massless graviton and the gravitino, whose
mass is given by the following formula derived in the leading approximation
\cite{W3}:
\begin{equation}
 \label{m-g}
      m_g^2=\frac{8\pi G}{3}\,\Lambda^4\ll \Lambda^2.
\end{equation}
To the low-energy approximation the heavy fields of inflatino and imaginary
part of scalar field do not propagate. The propagators and vertices of
interactions are presented in Appendix \ref{Feynman rules}. They are consistent with the superpotential and supercurrent of chiral superfield.

Then, the 1-loop terms in the effective potential of inflaton occur due to
the loops of
\begin{enumerate}
 \item the inflaton field itself,
 \item the gravitino from the supercurrent with the inflatino, which
     propagator is reduced to a constant at low energies,
 \item the graviton.
\end{enumerate}
In this section we analyze the model with the chiral superfield under the
regularization in the Euclidean space by means of momentum cut-off in the loop.

\subsection{Generation of $\mu_0$}
It is interesting to notice that the contraction of inflatino propagator into
the point at low energies due to Planckian scale of inflatino mass $m^\prime$
leads to a natural introduction of mass parameter in the superpotential,
$\mu_0$ initially equal to zero. Indeed, the vertexes for the coupling of
real scalar field to the inflatino due to the self-action constant $g_0$ and
for the interaction of inflatino with the gravitino due to the vacuum energy
at the scale $\Lambda$ are effectively reduced to the vertex for the coupling
of scalar field to the inflatino and gravitino (see fig.
\ref{fig-mu0})\footnote{We take into account for both the chiral rotation for
the inflatino (see Appendix \ref{kirrot}) and the Feynman rules (see Appendix
\ref{Feynman
 rules}).}, hence, one generates $\mu_0$ in the form
\begin{equation}
 \label{mu-0}
      \mu_0=2\,g_0\,\frac{\Lambda^2}{m^\prime}.
\end{equation}
Accounting for the scale of inflatino mass $m^\prime\sim {\tilde
m}_\mathrm{Pl}$, we straightforwardly see that the cosmological bootstrap
corresponds to the situation, when $g_0 \sim (\Lambda/{\tilde
m}_\mathrm{Pl})^2$ and, hence, $\mu_0\sim {\tilde m}_\mathrm{Pl}
(\Lambda/{\tilde m}_\mathrm{Pl})^4$, as it would be for the solution with the
set II.

\begin{figure}[th]\setlength{\unitlength}{0.85mm}
\begin{center}
 \begin{picture}(140,33)
 \put(0,0){\includegraphics[width=60\unitlength]{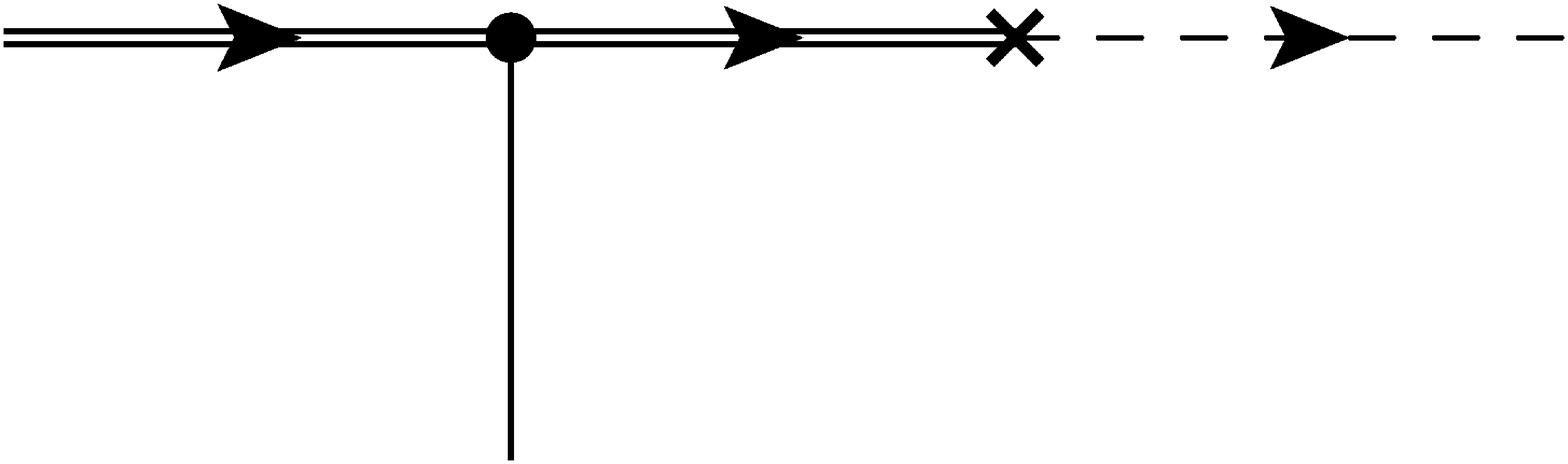}}
 \put(67,15){ $\boldsymbol\simeq$}
 \put(85,0){\includegraphics[width=40\unitlength]{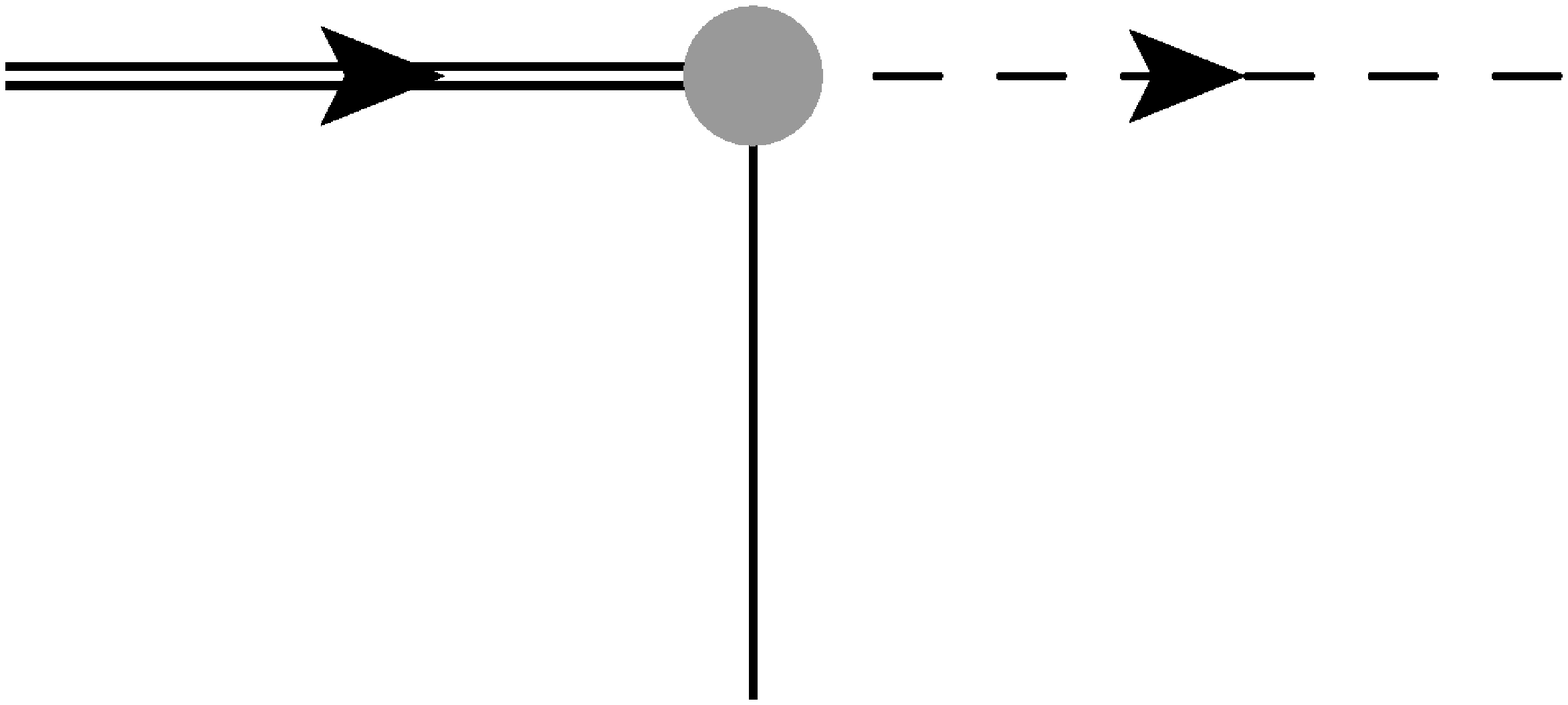}}
 \put(17,20){$g_0$}
 \put(37,20){$\Lambda^2$}
 \put(103,20){$\mu_0$}
\end{picture}
\end{center}
\caption{\label{fig-mu0}
The effective vertex for the coupling of inflaton to the inflatino and
gravitino after the contraction of inflatino propagator into the point;
the solid line denotes the propagator for the inflaton, the double line
does for the inflatino, the dashed line gives the gravitino; the vertex for
the coupling of inflatino to the inflaton is marked by the bold dot, while
the vertex for the transition of inflatino to the gravitino is denoted
by the cross.}
\end{figure}

Then, the initial bare superpotential with zero value of $\mu_0$ leads to the
potential $V_S^0=\Lambda^4(\hat\phi^2-1)^2$ [see (\ref{sponten-3})], i.e. it
corresponds to \textit{zero energy of vacuum}. Evidently, the generation of
effective mass parameter $\mu_0$ at low energy as described above occurs at
the tree level, hence it does not change the \textit{vacuum energy}, which
remains \textit{equal to zero}.

The variation of bare value (\ref{mu-0}) under the introduction of
corrections has got two following sources: $\delta g_0$ and $\delta
\Lambda^2$, so that
\begin{equation}\label{delta- mu-1}
      \delta \mu_0=\mu_0 \left(\frac{\delta g_0}{g_0}
      +\frac{\delta\Lambda^2}{\Lambda^2}\right)\approx
      \mu_0\,\frac{\delta\Lambda^2}{\Lambda^2},
\end{equation}
where the approximation has been accepted in accordance with the set II, when
the relative contribution by the variation of self-coupling constant $g_0$ is
suppressed. Therefore, following (\ref{shifts-prim}) and (\ref{delta- mu-1}),
we expect the correction of the form
\begin{equation}\label{delta-lambda-1}
      \frac{\delta\Lambda^2}{\Lambda^2}\sim -
      \left(\frac{\Lambda}{{\tilde m}_\mathrm{Pl}}\right)^2.
\end{equation}

Note,  in the case of cosmological bootstrap, relation (\ref{mu-0}) means
that there is the connection between the mass of inflatino with the primary
density of vacuum energy of Planckian scale, since the parameter $\mu_0$ in
set II is determined by the quantity $\omega_0$ [see (\ref{sets})] and its
value at the first step of bootstrap [see (\ref{cb-omega})], i.e. by the
cancelation of Planckian contributions to the vacuum energy, so that up to
small corrections we find
\begin{equation}\label{m-M}
      m^\prime=\frac{16}{5\sqrt{6}}\,M^2\sqrt{\pi\,G}\sim {\tilde m}_\mathrm{Pl}.
\end{equation}
This relation corresponds to the requirement of existing the flat potential
of module, and we do not consider this connection as the condition of any
``fine tuning'' for the model parameters, but, presumably, we put it as the
definition of the module itself, i.e. as the prerequisite of our model.

\subsection{Zero-point modes of field and regularization}
\begin{figure}[th]\setlength{\unitlength}{0.65mm}
\begin{center}
  \begin{picture}(40,40)
  \put(0,0){\includegraphics[width=35\unitlength]{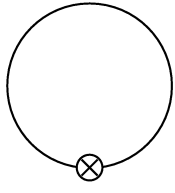}}
 \end{picture}
\end{center}
\caption{\label{fig-vac-loop} The vacuum loop of real scalar field with the
insertion of operator $p_\mu
p_\nu-\frac{1}{2}\,g_{\mu\nu}(p^2-m^2).$}
\end{figure}

A free real scalar-field has got the canonical tensor of energy-momentum
\begin{equation}\label{t-mn}
     T_{\mu\nu}=\partial_\mu\phi\,\partial_\nu\phi-
      \frac{1}{2}\,g_{\mu\nu}
      (\partial_\alpha\phi\,\partial^\alpha\phi- m^2\phi^2),
\end{equation}
and zero-point modes of field contribute due to the imaginary part of
propagator in the loop [see fig. \ref{fig-vac-loop}]:
\begin{equation}\label{vac-loop-Mink}
      \langle T_{\mu\nu}\rangle_0=\int\frac{\mathrm{d}^4p}{(2\pi)^4}\,
      \left\{p_\mu p_\nu-\frac{1}{2}\,g_{\mu\nu}(p^2-m^2)\right\}
      (\mathrm{i})^2\;
      \mathfrak{Im}\frac{1}{p^2-m^2+\mathrm{i}0}.
\end{equation}

Taking into account for
$$
      \mathfrak{Im}\frac{1}{p^2-m^2+\mathrm{i}0}=
      -\pi \delta(p^2-m^2),\qquad
      (p^2-m^2)\delta(p^2-m^2)=0,
$$
we find that the contribution to the vacuum expectation of energy-momentum
tensor due to the field lagrangian with the factor of metric
nullifies\footnote{Generally, this term represents the classical expression giving
the value of potential in its minimum, that is equal to zero for the free
field, of course.}, while in the Minkowskian space the formula
\begin{equation}\label{vac-loop-Mink2}
      \langle T_{\mu\nu}\rangle_0=\int\frac{\mathrm{d}^4p}{(2\pi)^4}\,
      p_\mu p_\nu \pi\delta(p^2-m^2)
\end{equation}
is reduced to the standard form for the zero-point modes after the
integration out of the delta-function over the temporal component of momentum
\begin{equation}\label{vac-loop-Mink3}
      \langle T_{\mu\nu}\rangle_0=\pi\int\frac{\mathrm{d}^3p}{(2\pi)^4
      2|p_0|}\,    p_\mu p_\nu\Big|_{p_0=\pm|p_0|} .
\end{equation}
This formula takes the sense under a regularization, for instance, due to the
cut off the spatial momentum by the scale $\Lambda_M$. Then, because of the
spherical symmetry the resulting averaged energy-momentum tensor corresponds
to an ultra-relativistic matter in the case, when the cut-off scale is
essentially greater than the field mass, $\Lambda_M\gg m$, or to a
non-relativistic matter, ``dust'', if the mass is essentially greater than
the cut-off $m\gg \Lambda_M$. Really,
$$
      \langle T_{00}\rangle_0=\int\frac{\mathrm{d}^3p}{(2\pi)^3}\,
      \frac{1}{2}|p_0|,       \qquad
      \langle T_{\alpha 0}\rangle_0=\langle T_{0\alpha}\rangle_0=0,
$$
$$
       \langle T_{\alpha\beta}\rangle_0=\int\frac{\mathrm{d}^3p}{(2\pi)^3
      2|p_0|}\,    p_\alpha p_\beta=\frac{1}{3}\,\delta_{\alpha\beta}
      \int\frac{\mathrm{d}^3p}{(2\pi)^3
      2|p_0|}\,p^2.
$$
However, such the result seems to be not physical, because such the
zero-point modes possess the energy-momentum tensor different from the vacuum
one, that should be proportional to the metric. This fact points to that a
regularization should be consistent with the physical requirements (see
\cite{ACGKS}).  The trivial condition would be the introduction of normal
order for operators in the definition of energy-momentum tensor, that simply
puts the contribution of zero-point modes to be equal to zero. We go in
another way, introducing the regularization in the Euclidean space after the
Wick rotation: $p_0=\mathrm{i}p_{4}$, so that the expression for the
contribution by the zero-point modes into the averaged tensor of
energy-momentum becomes equal to
\begin{equation}\label{vac-loop-Euclid}
      \langle T_{\mu\nu}\rangle_0^E=
      \int\frac{\mathrm{i}\,\mathrm{d}^4p_E}{(2\pi)^4}\,
      \frac{-\mathrm{i}}{p_E^2+m^2}\,
      p_\mu^E p_\nu^E=\frac{1}{4} g_{\mu\nu}^E
      \int\frac{\mathrm{d}^4p_E}{(2\pi)^4}\,
      \frac{p_E^2}{p_E^2+m^2}.
\end{equation}
where we have taken into account for the spherical symmetry of Euclidean
space. Thus, the contribution of single zero-point mode of real scalar field gives
\begin{equation}\label{vac-loop-E2}
      \langle T_{\mu\nu}\rangle_0^E=- g_{\mu\nu}
      \frac{1}{(16\pi)^2}\left\{\frac{1}{2}\,p_E^4
      -m^2p_E^2+m^4\ln\frac{p_E^2+m^2}{m^2}\right\}
      \Bigg|_{\Lambda_d^2}^{\Lambda_u^2},
\end{equation}
where $\Lambda_{u,d}$ denote the upper and lower boarders in the cut-off over
the Euclidean momentum, respectively.

For the Majorana fermion with the energy-momentum tensor
\begin{equation}\label{vac-loop-maj1}
      T_{\mu\nu}^\prime=\frac{1}{2}\,\bar\psi p_\mu\gamma_\nu \psi-
      \frac{1}{2}\,g_{\mu\nu}\,\bar\psi(\hat p-m^\prime)\psi,
\end{equation}
the analogous procedure taking into account for the elementary calculation of
trace for Dirac's gamma-matrices $\mbox{tr}[\gamma_\nu(\hat p+m^\prime)]=4
p_\nu$ and for the minus sign in the case of fermionic loop, leads to the
following expression for the contribution of zero-point modes:
\begin{equation}\label{vac-loop-maj2}
      \langle T_{\mu\nu}^\prime\rangle_0^E=g_{\mu\nu}
      \frac{2}{(16\pi)^2}\left\{\frac{1}{2}\,p_E^4
      -{m^\prime}^2p_E^2+{m^\prime}^4\ln\frac{p_E^2+
      {m^\prime}^2}{{m^\prime}^2}\right\}
      \Bigg|_{\Lambda_d^2}^{\Lambda_u^2},
\end{equation}
where factor 2 corresponds to two, left and right modes of Majorana particle
in comparison with the case of scalar field.

Summing up the contributions by components of chiral superfield into the
energy-momentum tensor of vacuum gives zero, if the exact supersymmetry takes
place (all of masses in the supermultiplet are equal to each other), while
after the breaking down the supersymmetry the sum rules
$$
      \sum  (-1)^F=0,\qquad \sum (-1)^F m^2=0,
$$
lead to the expression\footnote{Here $\tilde m$ denotes the mass of imaginary
part of scalar filed.}
\begin{equation}\label{sum-vac-loop}
      \langle T_{\mu\nu}\rangle^E=g_{\mu\nu}
      \frac{1}{(16\pi)^2}\left\{
      2{m^\prime}^4\ln\frac{p_E^2+{m^\prime}^2}{{m^\prime}^2}-
      \tilde m^4\ln\frac{p_E^2+\tilde m^2}{\tilde m^2}\right\}
      \Bigg|_{\Lambda_d^2}^{\Lambda_u^2},
\end{equation}
where we have neglected the contribution by the real part of scalar field,
because of approximating its mass equal to zero. Therefore, the sum rules
give $\tilde m^2=2 {m^\prime}^2$, and finally, we get
\begin{eqnarray}
      \langle T_{\mu\nu}\rangle^E &=& g_{\mu\nu}
      \frac{2}{(16\pi)^2}\left\{
      {m^\prime}^4\ln\frac{p_E^2+{m^\prime}^2}{{m^\prime}^2}-
      2{m^\prime}^4\ln\frac{p_E^2+2{m^\prime}^2}{2{m^\prime}^2}\right\}
      \Bigg|_{\Lambda_d^2}^{\Lambda_u^2},\nonumber\\[3mm]
      &=& g_{\mu\nu}
      \frac{2{m^\prime}^4}{(16\pi)^2}
      \ln\frac{1+\frac{p_E^2}{{m^\prime}^2}}
     {\left(1+\frac{p_E^2}{2{m^\prime}^2}\right)^2}\;
      \Bigg|_{\Lambda_d^2}^{\Lambda_u^2}. \label{sum-vac-loop-2}
\end{eqnarray}
Since the logarithm in formula (\ref{sum-vac-loop-2}) has got negative
values, we can draw the following conclusions:
\begin{itemize}
 \item if the upper limit of integration has got values of the order of
     Planckian scale, $\Lambda_u\sim {\tilde m}_\mathrm{Pl}$, then it
     gives the negative contribution into the density of vacuum energy,
     that corresponds to the cancelation of initial value of vacuum
     energy $M^4$ due to the introduction of suitable value of parameter
     $\omega_0$ in the superpotential;
 \item if the lower limit satisfies the condition $\Lambda_d\ll
     m^\prime$, then we get
\begin{equation}\label{lim-vac}
       \langle T_{\mu\nu}\rangle^E_u = g_{\mu\nu}\,
      \frac{1}{(16\pi)^2}\,
      \frac{1}{2}\,\Lambda_d^4\;\sim \;
      g_{\mu\nu}\,\Lambda^4,
\end{equation}
that corresponds to the introduction of vacuum energy of the order of
$\Lambda^4$ at $\Lambda_d\sim \Lambda$.
\end{itemize}
Thus, the integration over the Euclidean momentum squared in limits
$[\Lambda_E^2,M_E^2]$ at $\Lambda_E\sim \Lambda$ and $M_E\sim {\tilde
m}_\mathrm{Pl}$ forms the initial conditions for the introduction of chiral
superfield of module. The interval of integration $[0,\Lambda_E]$, i.e. the
regularization in the Euclidean space with the cut-off $\Lambda_E$,
corresponds to the contribution at low energies due to the light fields. In
what follows, we will consider similar intervals of integrations in the
calculation of loops for the effective potential.

Then, the correction to the energy-momentum tensor of vacuum due to the
zero-point modes gets the form\footnote{This correction certainly corresponds
to the leading contribution of real scalar field being the only component of
chiral superfield that possesses the mass essentially less than the cut-off;
at such the cut-off  the ``heavy'' field components give small corrections to
$\delta T_{\mu\nu}^E$ of the order of $(\Lambda/{\tilde m}_\mathrm{Pl})^2$.}
\begin{equation}\label{vac-E}
      \delta T^E_{\mu\nu}=-g_{\mu\nu}\,
      \frac{1}{(16\pi)^2}\,
      \frac{1}{2}\,\Lambda_E^4.
\end{equation}

It is important to note that the separation of Planckian interval of
integration with the introduction of intermediate cut-off $\Lambda_E$ as
given above, implies that we get the exact cancelation of vacuum energy of
the order of $\Lambda^4$! Moreover, we clearly see that the switching on the
field interactions with the coupling constants in the form of power
corrections in $\Lambda/{\tilde m}_\mathrm{Pl}$ does not change our
statements.

\subsection{Generation of $g_0$}

\begin{figure}[th]\setlength{\unitlength}{0.75mm}
 \begin{center}
  \begin{picture}(70,20)
   \put(0,0){\includegraphics[width=60\unitlength]{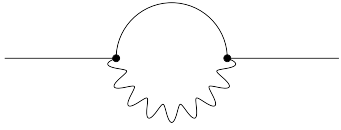}}
  \end{picture}
 \end{center}
\caption{\label{fig-grav-g} The gravitational correction to the quadratic
term in the effective potential of inflaton (the wave line denotes the graviton).}
\end{figure}

While the origin of mass parameter $\mu_0$ can be quite naturally explained
by the introduction of effective constant due to the contraction of inflatino
propagator in the case of non-zero self-action of the field, the appearance
of self-action itself with the coupling constant $g_0$, corresponding to the
cosmological bootstrap, is not trivial, because it is related to the
existence of primary module field. However, we can clarify this issue by the
following note: the modification of canonical energy-momentum tensor of free
real scalar field by the formula\footnote{The momentum operator acts to one
of scalar fields, for instance, to the right one. Of course, we can write
down the expression in terms of partial derivatives, which symmetrically act
twice to the field in the right and to the field in the left.}
\begin{equation}\label{mod-T}
      T_{\mu\nu}^\mathrm{mod}=\phi\, p_\mu p_\nu\,\phi-
      \frac{1}{2}\,g_{\mu\nu} \phi(p^2-m^2)\phi,
\end{equation}
due to the account for the interaction with the graviton, leads to quadratic
term of scalar field interaction in the effective potential (see the diagram
in fig. \ref{fig-grav-g}).

\noindent Indeed, the calculation of loop in the Euclidean space up to the
leading approximation\footnote{Here, we neglect the contribution by the mass
term.} gives
\begin{equation}\label{lopp-g0}
      -\mathrm{i}\,2g_0\Lambda^2=
      \int\frac{\mathrm{i}\,\mathrm{d}^4p_E}{(2\pi)^4}\,
      {4\pi G}\quad\Rightarrow\quad
      g_0=-\frac{G}{16\pi}\,\frac{c_g\Lambda_E^4}{\Lambda^2},
\end{equation}
where we have introduced the dimensionless constant $c_{g}$ in order to
parameterize the arbitrary choice of ultraviolate cut-off in agreement to the
procedure of renormalizations. In this way, we suggest that $c_g$ is of the
order of unit. Therefore, first, the generated value of bare constant $g_0$
has got the desired order in the ratio $\Lambda/{\tilde m}_\mathrm{Pl}$, and
second, it can get the required value for the leading approximation in the
bootstrap, if we put the cut-off equal to
\begin{equation}\label{g_0-bare}
      c_g\Lambda_E^4=\frac{2}{3}\,(16\pi\Lambda^2)^2\quad\Rightarrow
      \quad
      g_0=-\frac{32\pi G}{3}\,\Lambda^2.
\end{equation}
In this case, the correction to the vacuum energy due to the loop of
zero-point modes has got the form
\begin{equation}\label{vac-E2}
      \delta T^E_{\mu\nu}=-\frac{1}{3c_g}\,g_{\mu\nu}\,\Lambda^4,
\end{equation}
that show the necessity of contribution by other fields into the vacuum energy
of the order of $\Lambda^4$, if $c_g\neq \frac13$. In this respect, we have
to note that we have suggested that the loop under the consideration gives
the contribution to the constant of quadratic self-action of field, while the
field normalization remains unchanged, though generically one has to take
into account for the opportunity of such the renormalization in powers of
$\Lambda/{\tilde m}_\mathrm{Pl}$.

The above study shows that primary parameters of model can be generated due
to the introduction of loop corrections with virtual gravitons and, hence,
gravitinos to the free scalar field, if the cut-off is set by the scale of
primary cosmological constant $\Lambda$, by the order of magnitude, of
course. However, it means that loops with gravitons and gravitinos should not
be taken into account with the same cut-off, while we consider the breaking
of cosmological bootstrap, or we have to put the corresponding cut-off in
loops with gravitons and gravitinos to be suppressed by (even) degrees in
$\Lambda/{\tilde m}_\mathrm{Pl}$, otherwise we would not distinguish the
leading order from the corrections. The same conclusion can be drawn, if we
consider the contributions of zero-point modes of graviton and gravitinos
into the vacuum energy: at the cut-off taken of the order of $\Lambda$ such
the modes occur inadmissibly large, i.e. they get the same order of
$\Lambda^4$ as the matter fields, although we would expect that the
gravitational field contribute as corrections to the vacuum energy suppressed
by the small ratio $(\Lambda/{\tilde m}_\mathrm{Pl})^2$, at least. Further we
will convince that the model of cosmological bootstrap would be reasonable,
if only one-loop contributions by gravitons and gravitinos have got the
cut-offs with the mentioned suppression in $\Lambda/{\tilde m}_\mathrm{Pl}$.

\subsection{Inflaton loops}

\begin{figure}[th]
 \begin{center}
  \includegraphics {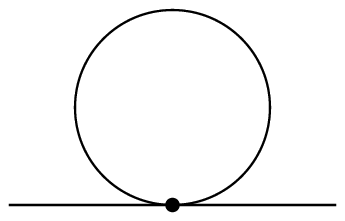}\\[3mm]
  \includegraphics {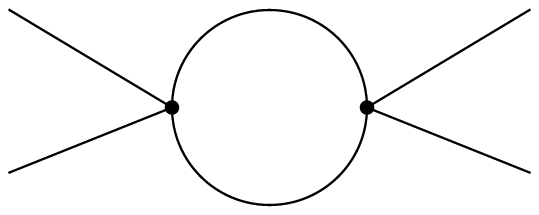}
 \end{center}
\caption{\label{fig-2legs} The 1-loop corrections to the effective potential
of inflaton: the self-action of second and fourth degrees in the field.}
\end{figure}

The one-loop corrections caused by the real scalar field only, corresponds to
contributions into the effective potential as shown in diagrams represented
in fig. \ref{fig-2legs}, 
i.e. they give the corrections to the field self-action with degrees of 2 and 4.
The amplitudes are given by expressions
\begin{equation}\label{L1}
-\mathrm{i}L_1=-3\,\frac{g_0^2}{16\pi^2}\,\mathrm{i}\,c_2\Lambda_E^2,\qquad
      -\mathrm{i}L_2=54\,\frac{g_0^4}{16\pi^2}\,\mathrm{i}\,
      \ln\frac{c_4\Lambda_E^2}{\Lambda_\mathrm{reg}^2},
\end{equation}
corresponding to the corrections to the lagrangian
$$
      \delta_1\mathcal{L}=\frac{1}{2}\,L_1\,\phi^2+\frac{1}{4!}\,L_2\,\phi^4,
$$
whereas in (\ref{L1}) we have introduced the normalization scale of
logarithmic corrections $\Lambda_\mathrm{reg}$ as well as the dimensionless
constants $c_{2,4}$, corresponding to variations of cut-off parameter for
different physical quantities in the regularization and renormalization.

Then, $L_2$ provides us with the standard renormalization of constant $g_0$:
\begin{equation}\label{RG-g}
      \delta g_0=-9\,\frac{g_0^3}{32\pi^2}\,
      \ln\frac{c_4\Lambda_E^2}{\Lambda_\mathrm{reg}^2}
      \quad\Rightarrow\quad
      \frac{\mathrm{d}g_0}{\mathrm{d}\ln\Lambda_\mathrm{reg}}=9\,
      \frac{g_0^3}{16\pi^2},
\end{equation}
so that at $\Lambda_\mathrm{reg}<\Lambda_E$ and $g_0<0$ it leads to
$$
      \delta g_0\sim\left(\frac{\Lambda}{{\tilde m}_\mathrm{Pl}}
      \right)^6>0,
$$
as we could expect in the bootstrap model, set II (see (\ref{shifts-prim})).

Further, $L_1$ can be naturally treated as the correction to the primary
value of $2g_0\Lambda^2$, determining the quadratic self-action of field,
whereas, because the variation of $g_0$ has got the large degree in small
${\Lambda}/{{\tilde m}_\mathrm{Pl}}$ as found, we have to put
$$
      L_1=2g_0\delta\Lambda^2,
$$
resulting in
\begin{equation}\label{RG-l}
      \delta \Lambda^2 =\frac{3}{2}\,\frac{g_0}{16\pi^2}\,c_2\Lambda_E^2<0.
\end{equation}
In accordance to (\ref{delta- mu-1}) we get
\begin{equation}\label{delta-mu-2}
     \frac{\delta\mu_0}{\mu_0}=\frac{3}{2}\,\frac{g_0}{16\pi^2}\,
      \frac{c_2\Lambda_E^2}{\Lambda^2}\sim -
      \left(\frac{\Lambda}{{\tilde m}_\mathrm{Pl}}
      \right)^2<0.
\end{equation}

Let us redefine the dimensionless quantities with hats in (\ref{hats}) by the
substitution of $\Lambda^2\mapsto \Lambda^2+\delta\Lambda^2$. Such the
procedure gives the potential of scalar field in the form
\begin{equation}\label{cb-poten}
      V=(\Lambda^2+\delta\Lambda^2)^2\left(
      1-\frac{\hat\mu^2}{2}\,\hat\phi^2+\frac{\hat\lambda}{4}\,\hat\phi^4
      \right),
\end{equation}
where
\begin{equation}\label{cb-param}
 \begin{array}{rcccl}
      \hat\mu^2&=&\displaystyle
      -5\frac{\delta\mu_0}{\mu_0}&=&\displaystyle
      -\frac{15}{2}\,\frac{g_0}{16\pi^2}\,
      \frac{c_2\Lambda_E^2}{\Lambda^2},\\[5mm]
      \hat\lambda&=&\displaystyle
      4\,\frac{\delta g_0}{g_0}&=&\displaystyle
      -9\,\frac{g_0^2}{4\pi^2}\,
      \ln\frac{\Lambda_E\sqrt{c_4}}{\Lambda_\mathrm{reg}},
 \end{array}
\end{equation}
and we evidently see that we can easily chose a reasonable value of
renormalization point $\Lambda_\mathrm{reg}$, which would give zero value of
vacuum energy density: $\hat\mu^4=4\hat\lambda$. At this condition, the
normalization point becomes close to the vacuum expectation value of field
itself;
\begin{equation}\label{vev}
      \langle \phi\rangle^2=\frac{3}{16\pi G}\,\langle \hat\phi\rangle^2=
      \frac{3}{4\pi G\hat\mu^2}\sim\Lambda_\mathrm{reg}^2.
\end{equation}

Thus, the considered contributions provide us with the natural realization of
cosmological bootstrap.

However, loops of scalar field also have lead to corrections for vertexes of
contact interaction between the inflaton, gravitino and inflatino due to the
introduction of supercurrent, so that these loops are analogous to those of
calculated above. In addition, we have to emphasize that at low energies
inflatino with the mass of Planckian scale is ``frozen'', hence, external
inflatino fields are equal to zero, and therefore, loops of scalar fields in
the mentioned vertexes of supercurrent have no relation to the considered
contributions, because external inflatino-legs in the diagrams lead to zero
values of those corrections.

\subsection{Loops of gravitino and gravitons}

To find the correction quadratic in the inflaton field let us consider the
loop diagram\footnote{In this section all of integrals are calculated in the
Euclidean space.} with the gravitino and inflatino (fig. \ref{fi2}) in
accordance with the Feynman rules (see Appendix \ref{Feynman rules}),
\begin{figure}[ht]
 \begin{center}
{\centering{\includegraphics {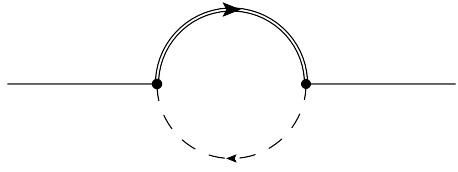}}}
\end{center}
\caption{The contribution of loop with the inflatino
into the quadratic self-action of inflaton.}
\label{fi2}
\end{figure}
wherein we put the incoming and outcoming momenta of inflaton to be equal to
zero, while the loop momentum is denoted by $k$. Then
\begin{eqnarray}\label{mu2}
\delta \mu^2&=& -\frac{\mu_0^2 (8\pi G)}{4} \int \frac{\ud ^4
k}{(2\pi)^4} Tr \left\{\frac{P(k)^{\mu\nu}}{k^2+m_g^2} \gamma_\nu
\gamma^5 \frac{-i\gamma k+m'}{k^2+m'^2} \gamma_\mu \gamma^5
\right\}\\ \nonumber &=&-\frac{(8\pi G)\mu_0^2 c_m \Lambda_E^4}{24\pi^2
m_g}\left( \frac1{m'} +\mathcal{O}\left(\frac1{m'^2}\right) \right).
\end{eqnarray}
Here we introduce a constant $c_m$ as the parameter of ultraviolate cut-off.
Since the gravitino propagate in the loop, we have to put $c_m\ll1$, so that
taking into account for the expression giving the gravitino mass, we get that
the contribution under study is suppressed as
$$
      \frac{\delta \mu^2}{\mu_0^2}\sim -c_m
      \left(\frac{\Lambda}{{\tilde m}_\mathrm{Pl}}\right)^2,
$$
and it is not essential in respect of consideration for the breaking down the
cosmological bootstrap.

\begin{figure}[ht]
\begin{center}
{\centering{\includegraphics {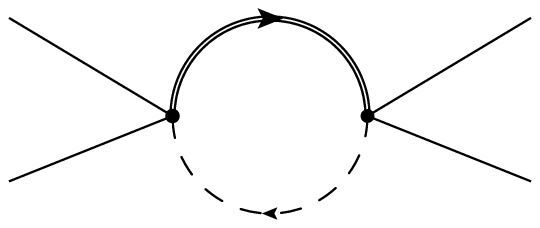}}}
\end{center}
\begin{center}
{\centering{\includegraphics {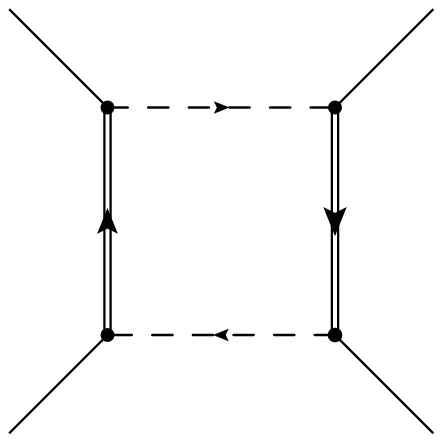}}}
\end{center}
\caption{The contribution of loops with the inflatino into the quartic self-action
of inflaton.}
\label{fi4}
\end{figure}

The correction to $g_0$ are given by two diagrams with zero outcoming momenta
and the loop momentum $k$ (fig. \ref{fi4}).

In accordance with the first diagram in fig. \ref{fi4} we get
\begin{eqnarray}\label{69}
g_0\delta g_0&=& \frac{3}{12} \frac{g_0^2 (8\pi G)}{2} \int \frac{\ud
^4 k}{(2\pi)^4} Tr \left\{\frac{P(k)^{\mu\nu}}{k^2+m_g^2}
\gamma_\nu
 \frac{-i\gamma k+m'}{k^2+m'^2} \gamma_\mu
\right\}\\ \nonumber&=& -\frac{(8\pi G) g_0^2 c_g^\prime\Lambda_E^4}{48\pi^2
m_g}\left( \frac1{m'} +\mathcal{O}\left(\frac1{m'^2}\right) \right)
\sim -g_0^2 c_g^\prime\left(\frac{\Lambda}{{\tilde m}_\mathrm{Pl}}\right)^2.
\end{eqnarray}
The second diagram in fig. \ref{fi4} results in the following analytic
expression:
\begin{eqnarray}\label{70}
g_0\delta g_0& =& \frac{6}{12} \frac{\mu_0^4 (8\pi G)^2}{16}
\int
\frac{\ud ^4 k}{(2\pi)^4}\times\nonumber\\ && Tr
\left\{\frac{P(k)^{\mu\nu}}{k^2+m_g^2} \gamma_\nu \gamma^5
\frac{-i\gamma k+m'}{k^2+m'^2} \gamma_\sigma \gamma^5
\frac{P(k)^{\sigma\rho}}{k^2+m_g^2} \gamma_\rho \gamma^5
\frac{-i\gamma k+m'}{k^2+m'^2} \gamma_\mu \gamma^5 \right\}
\\
\nonumber &\approx &\frac{\mu_0^4 (8\pi G)^2 c_g^{\prime\prime}\Lambda_E^2}{8
m'^2}+\mathcal{O}\left(\frac{1}{m'^3}\right)
\sim -g_0 c_g^{\prime\prime}\left(\frac{\Lambda}{{\tilde m}_\mathrm{Pl}}\right)^{12}.
\end{eqnarray}
From (\ref{69}) we see that, in the presence of gravitino within the loop,
the constant of cut-off should be suppressed as $c_g^\prime\sim
({\Lambda}/{{\tilde m}_\mathrm{Pl}})^2$. Then, the model of cosmological
bootstrap remains consistent, though the inflatino gives an essential
contribution into variation of field self-action as large as comparable to
the logarithmic renormalization, and probably, it plays a dominant role.
Nevertheless, this fact does not change estimates in the order of magnitude
as was done in previous sections. From (\ref{70})  we see that the role of
constant $c_g^{\prime\prime}$ is inessential.

\begin{figure}[ht]
\begin{center}
\includegraphics {gravit1.eps}
\end{center}
\begin{center}
\includegraphics {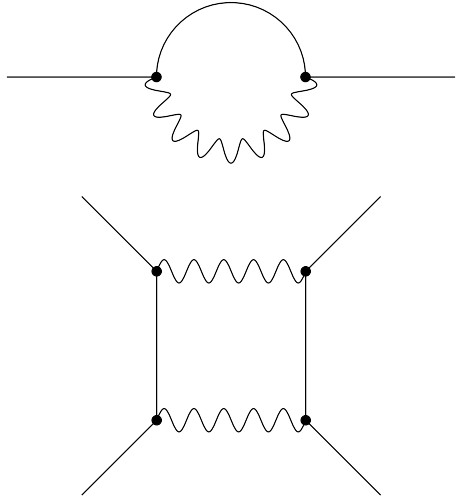}
\end{center}
\caption{The contribution of loops with gravitons (waved line) into the
self-action of inflaton.}
\label{figra}
\end{figure}

Corrections caused by the interaction of canonical tensor of energy-momentum
with the graviton are given by diagrams in fig. \ref{figra}.

The first diagram in fig. \ref{figra} gives the logarithmic correction to the
mass in the form of
\begin{eqnarray}
\delta \mu^2&=& -(8\pi G) (\mu_0^2+2g_0 \Lambda^2)^2 \eta_{\mu\nu} (
\eta^{\mu\mu'}\eta^{\nu\nu'}+\eta^{\mu\nu'}\eta^{\nu\mu'}-\eta^{\mu\nu}\eta^{\mu\nu'})
\eta_{\mu'\nu'}\times\nonumber
\\&& \int \frac{\ud^4 k }{(2\pi)^4} \frac1{2k^2}
\frac1{k^2} 
=\frac{(8\pi G) (\mu_0^2+2g_0
\Lambda^2)^2}{4\pi^2}\ln\frac{c_m^\prime\Lambda_E^2}{\Lambda_{\mathrm{reg}}^2}
\label{71}
\\
&
\sim& \Lambda^2 \left(\frac{\Lambda}{{\tilde m}_\mathrm{Pl}}\right)^6
\ln\frac{c_m^\prime\Lambda_E^2}{\Lambda_{\mathrm{reg}}^2},
\nonumber
\end{eqnarray}
while the second diagram in fig. \ref{figra} does the correction to the
constant of self-action
\begin{eqnarray}
g_0\delta g_0&=&-\frac{6}{12} (8\pi G)^2 (\mu_0^2+2g_0 \Lambda^2)^4
\eta_{\mu\nu} (
\eta^{\mu\mu'}\eta^{\nu\nu'}+\eta^{\mu\nu'}\eta^{\nu\mu'}-\eta^{\mu\nu}\eta^{\mu\nu'})
\eta_{\mu'\nu'}\times\nonumber\\\label{72}
&&\eta_{\mu_1\nu_1} (
\eta^{\mu_1\mu'_1}\eta^{\nu_1\nu'_1}+\eta^{\mu_1\nu'_1}\eta^{\nu_1\mu'_1}-\eta^{\mu_1\nu_1}\eta^{\mu_1\nu'_1})
\eta_{\mu'_1\nu'_1} \int \frac{\ud^4 k }{(2\pi)^4} \frac1{4k^4}
\frac1{(k^2)^2}\\ \nonumber&=&\frac{(8\pi G)^2 (\mu_0^2+2g_0
\Lambda^2)^4}{4\pi^2}\left\{
\frac1{\tilde c_g\Lambda_E^4}-\frac1{\Lambda^4_{\mathrm{reg}}}\right\}
\sim -g_0 \,\frac{1}{\tilde c_g}\left(\frac{\Lambda}{{\tilde m}_\mathrm{Pl}}\right)^{10}.
\end{eqnarray}
Contribution (\ref{71}) can be treated as the suppressed logarithmic
correction to the scale $\Lambda^2$ like $g_0\delta\Lambda^2$, that is
inessential, or as the variation of $\delta g_0$, which can be comparable
with the logarithmic correction due to the loop with the inflaton calculated
above.  In the second case, in accordance with $c_m^\prime\sim
{\Lambda}/{{\tilde m}_\mathrm{Pl}}$
we find that the logarithm argument has got an additional power in the small
parameter, though this fact makes it value to be of the same order as the
renormalization correction. In this way, the correction in (\ref{71}) has got
the opposite sign, so that one can get a compensation of contributions.
Probably, this fact points that the main contribution is given by
(\ref{69}).

Following (\ref{72}) we conclude that even at $\tilde
c_g\sim({\Lambda}/{{\tilde m}_\mathrm{Pl}})^2$ this contribution is not
essential for our consideration.

Thus, we have shown that the one-loop structure of theory is consistent with
relations required for the reasonable breaking down the cosmological
bootstrap. It means that the instability of primary cosmological constant is
matched to the inflation of Universe.

\section{Conclusion}

We have confirmed the possibility to construct the realistic model, in which
\begin{itemize}
 \item the primary cosmological constant with the characteristic scale of
     the order of grand unification energy corresponds to the flat
     potential of real scalar field-module, that possesses the
     non-trivial superpotential, whose contribution is modified after the
     account for the leading corrections in gravitational constant within
     the supergravity,
\item the fine tuning of superpotential parameters should be investigated
    with respect of stability, and it is broken due to quantum
    loop-corrections, that leads to the instability of primary
    cosmological constant,
\item the primary cosmological constant relaxes due to the inflation
    caused by the instability of potential for the module field playing
    the role of inflaton,
\item after the account for quantum loop-corrections, the parameters of
    potential naturally agree with the values phenomenologically observed
    in the description of large scale structure of Universe, by the order
    of magnitude.
\end{itemize}
The existence of primary superpotential for the module field is determined by
the relation between its parameters, which is called the cosmological
bootstrap.

In the model offered, there are only two significant parameters being the
scales of energies: the Planckian mass ${\tilde m}_\mathrm{Pl}$ and the scale
of primary cosmological constant of the order of energy in the grand
unification $\Lambda$ constrained by the strict hierarchy $\Lambda/{\tilde
m}_\mathrm{Pl}\ll 1$. This proposition is enough for the realization of
cosmological bootstrap as well as for the natural explanation of inflaton
parameters.

In the second part of paper we have performed calculations of 1-loop
structure of theory. In this way we have found the conditions allowing for
the necessary hierarchy of loop corrections suitable for the cosmological
bootstrap in order to break down the fine tuning of primary superpotential: the
cut-off in loops with the inflaton should be about $\Lambda$, while in loops
with graviton, gravitino and inflatino one should suppress the cut-off by the
factor of $(\Lambda/{\tilde m}_\mathrm{Pl})^2$.

We have to note several technical issues: first, we have considered the
corrections to the potential due to the supergravity in the leading order in
the constant of gravitational interaction. Second, we have taken into account
for contributions into the renormalizable terms of scalar field self-action,
i.e. into the self-action below the quartic one. In this way, we have assumed
that the self-action of higher powers in inflaton is beyond the control,
because such the terms of lagrangian permit an arbitrary final
renormalization, therefore, we have assumed them to be negligibly small in
our model. Third, the scheme with the fine tuning of potential parameters,
i.e. the cosmological bootstrap, includes the fixed expansion of parameters
in higher degrees of small ratio of cosmological scale to the Planckian mass
$\Lambda/{\tilde m}_\mathrm{Pl}$. Therefore, the introduction of loop
corrections in $\Lambda/{\tilde m}_\mathrm{Pl}$ breaks the structure of
cosmological bootstrap and generates the conditions of inflation. In this
respect, the offered scheme for the relation of primary cosmological constant
to the inflation looks like reasonable, in a whole.


This work is in part supported by the RFBR grants 10-02-00061, the grant of
Russian Federal Program ``Science and Education personal'' for the Center of
Science and Education 2009-1.1-125-055-008; the work by S.A.T. is supported
by the grant of Russian President MK-406.2010.2.

\appendix
\def\thesection{\Roman{section}}
\section{\label{appen} An imaginary part of scalar field and a general form of
potential} With no restriction of generality in the consideration of
superpotential (\ref{spot-x}) one can put the parameter $\Lambda$ to be real,
while $\mu_0$ and $g_0$ should be generically set complex. Then, the
potential gets the form
\begin{equation}\label{ap1}
    \begin{array}{ccl}\displaystyle
      V_S=\left|\frac{\partial W}{\partial\Phi}\right|^2&=&\displaystyle
      \Lambda^4+g_0g_0^*(\Phi^*\Phi)^2+
      \mu_0\mu_0^*\Phi^*\Phi 
      +\Lambda^2\big\{g_0^*(\Phi^*)^2+g_0\Phi^2\big\}\\[2mm]
      &+&\mathrm{i}\big\{\mu_0\Phi-\mu_0^*\Phi^*\big\}+
      \mathrm{i}\Lambda^2\Phi^*\Phi\big\{\mu_0g_0^*\Phi^*-\mu_0^*g_0\Phi\big\},
    \end{array}
\end{equation}
so that the discrete symmetry with respect to the inversion
$\Phi\leftrightarrow -\Phi^*$ leads to the following conditions:
\begin{equation}\label{ap2}
    \mu_0^*=\mu_0,\qquad g_0^*=g_0,
\end{equation}
and the potential equals
\begin{equation}\label{ap3}
      V_S=\Lambda^4+
      \frac{1}{2}\big\{\mu_0^2+2g_0\Lambda^2\big\}\phi^2+\frac{1}{4}\,g_0^2\phi^4+\Delta V_S,
\end{equation}
where the additional term is equal to
\begin{equation}\label{ap4}
    \Delta
    V_S=\mu_0\big\{g_0\phi^2-2\Lambda^2\big\}\tilde\phi-4g_0\Lambda^2\tilde\phi^2,
\end{equation}
as expressed in real fields
\begin{equation}\label{ap5}
    \phi=\sqrt{2}|\Phi|,\qquad \tilde\phi=\Im\mathfrak{m}\,\Phi,
\end{equation}
hence, there is the constraint
\begin{equation}\label{ap6}
    |\tilde \phi|\leqslant\frac{1}{\sqrt{2}}\,\phi.
\end{equation}
In the model under study we have got $g_0<0$, therefore, the term $\Delta
V_S$ has got the minimum versus $\tilde\phi$ at a fixed absolute value of $\Phi$.
In this way, the nil value of imaginary part for the field is not stable.

The minimum versus $\tilde\phi$ is posed at
\begin{equation}\label{ap7}
    \tilde\phi_\star=\mu_0\,\frac{g_0\phi^2-2\Lambda^2}{8g_0\Lambda^2},
\end{equation}
hence, this field can be approximated by a constant
\begin{equation}\label{ap8}
    \tilde\phi_\star\approx -\frac{\mu_0}{4g_0},
\end{equation}
in the region of
\begin{equation}\label{ap9}
    \phi^2\ll -\frac{\Lambda^2}{g_0},
\end{equation}
if $\phi^2>\mu_0^2/8 g_0^2$. In the realistic model we get $g_0\sim -
(\Lambda/{\tilde m}_\mathrm{Pl})^6$, and the constraint of (\ref{ap9}) is
reduced to $\phi\ll {\tilde m}_\mathrm{Pl} ({\tilde
m}_\mathrm{Pl}/\Lambda)^2$, i.e. to the following condition: the field is
certainly less than its vacuum expectation value.

The substitution of (\ref{ap7}) gives the potential for the field $\phi$,
\begin{equation}\label{ap10}
    V_S\mapsto \frac{\lambda_0}{4}\left(\phi^2-\phi_0^2\right)^2,
\end{equation}
where
\begin{equation}\label{ap11}
    \phi_0^2=-\frac{2\Lambda^2}{g_0},\qquad
    \lambda_0=g_0\left(g_0+\frac{\mu_0^2}{4\Lambda^2}\right),
\end{equation}
i.e. the potential has got zero value of vacuum energy as should be in the
case of single chiral superfield, when one can find a suitable complex
solution of quadratic equation $\partial W/\partial\Phi=0$ with respect to
$\Phi$, that coincides with $\phi_0=|\Phi|$, of course.

Thus, we have draw the conclusion that in the case of actual cosmological
role for the scalar field, the phenomenological requirement on ``freezing
out'' its imaginary part in the dynamics means the introduction of term
breaking down the supersymmetry, that leads to $\tilde\phi \to 0$. We can
reach this purpose, for instance, by adding ``a massive term'' in the form
\begin{equation}\label{ap12}
    \Delta \tilde V=\tilde m^2\left(\tilde\phi+C\,\mu_0\frac{\Lambda^2}{\tilde m^2}\right)^2+\mbox{const.},
\end{equation}
where the mass has got a value of the order of Planck scale of energy,
$\tilde m\sim {\tilde m}_\mathrm{Pl}$. The first variant of $\Delta \tilde V$
with $C=0$ does not involve any fine tuning and it gives a negligible value
of imaginary part for the scalar field
$\tilde\phi_\star\sim\mu_0\Lambda^2/{\tilde m}_\mathrm{Pl}^2\to 0$, that
weakly depends on $\phi$, while the second variant with $C=1$ leads to the
cancelation of term linear in $\tilde \phi$ in the potential independent of
the real part of field, hence, one gets the stronger suppression of imaginary
part. The second variant essentially extends the region of applicability for
the approximation with zero value of $\tilde\phi$. Finally, in the third
variant at $C=2+\mu_0^2/(2g_0\Lambda^2)$ the imaginary part becomes equal to
zero in the vacuum, i.e. in the minimum of potential for the real part of
field at $\phi_\star^2=-2\Lambda^2/g_0-\mu_0^2/g_0^2>0$, then, in this
phenomenologically actual case, the vacuum state becomes invariant with
respect to the operation of complex conjugation for $\Phi$. In this way, the
dynamics of imaginary part is absolutely inessential at the energies much
less than the Planckian mass, i.e. in the classical description of gravity.

Thus, the involvement of scalar field in the cosmological model allows us to
assign this field to real values, indeed.

Finally, let us note that the introduction of mass for the imaginary part of
scalar field along with the breaking down the supersymmetry leads to the
standard sum rules for the squares of masses for the components of chiral
superfield with the fermionic number $F=\{0,1\}$
\begin{equation}
 \label{sumrule-m}
      \sum(-1)^F m^2=0\quad \Rightarrow \quad
      m^2+\tilde m^2=2{m^\prime}^2,
\end{equation}
where $m^\prime$ denotes the mass of scalar field superpartner being the
Majorana field of inflatino. Therefore, the inflatino inevitably gets the
mass of the order of Planck scale.

\section{A chiral rotation \label{kirrot}}
The superpartner of scalar field, i.e. the inflatino, is the Majorana spinor
$\psi$. It is the charge self-conjugated spinor possessing the left-handed
and right-handed components in the chiral representation
\begin{equation}  \label{maiorana-1}
      \psi=
      \left(\begin{array}{c}
      \psi_L\\
      \psi_R
           \end{array}
      \right)=
      \left(\begin{array}{c}
      \chi\\
      \bar\chi
           \end{array}
      \right),
\end{equation}
where the 2-component spinors are charge conjugated to each other,
$\bar\chi=\mathrm{i}\sigma_2\chi^*$, so that the superpotential
(\ref{spot-x}) leads to the terms of lagrangian quadratic in the inflatino
\begin{equation} \label{quadro}
      \mathcal{L}_2=-\frac{1}{2}\left(\mathrm{i}\mu_0\chi\chi-\mathrm{i}
      \mu_0\bar\chi\bar\chi+
      2g_0\Phi\chi\chi+2g_0\Phi^*\bar\chi\bar\chi\right).
\end{equation}
This lagrangian can be reduced to the standard form with the real mass of
Majorana field, if we make the chiral rotation
\begin{equation}\label{chiral-1}
      \psi_u=\mathrm{e}^{\mathrm{i}\gamma_5u}\psi\quad\Rightarrow\quad
      \chi_u=\mathrm{e}^{\mathrm{i}u}\chi,\quad
      \bar\chi_u=\mathrm{e}^{-\mathrm{i}u}\bar\chi,
\end{equation}
with the parameter $u=-\pi/4$, so that
\begin{eqnarray} \label{quadro-2}
     \mathcal{L}_2&=&-\frac{1}{2}\left(\mu_0\chi_u\chi_u+
      \mu_0\bar\chi_u\bar\chi_u-
   2\mathrm{i}g_0\Phi\chi_u\chi_u+
      2\mathrm{i}g_0\Phi^*\bar\chi_u\bar\chi_u\right)\\[ 3mm]
      &=&-\frac{1}{2}\left(\mu_0\bar\psi_R\psi_L+\mu_0\bar\psi_L\psi_R-
      2\mathrm{i}g_0\Phi\bar\psi_R\psi_L+
      2\mathrm{i}g_0\Phi^*\bar\psi_L\psi_R\right)\\[3mm]
      &=&-\frac{1}{2}\left(\mu_0\bar\psi\psi-
      \sqrt{2}\,\mathrm{i}\,g_0\phi\bar\psi
      \gamma_5\psi+2g_0\tilde\phi\bar\psi \gamma_5\psi\right),
\end{eqnarray}
where, for the sake of brevity of notations, we have omitted the subscript
$u$, that is not essential for our presentation.

Thus, after the chiral rotation we have got the Majorana field with the
definite vertexes of Yukawa interaction with the real and imaginary parts of
scalar field in the chiral superfield. In this respect, it is important to
emphasize that the kinetic terms for the left-handed and right-handed
components of spinor remain invariant under the chiral rotation. Moreover,
the contact terms of inflatino coupling to the gravitino and scalar field are
also invariant, that is caused by the fact that the gravitino is described by
the Majorana field, too, and the contact terms of interaction preserve the
chirality.

\section{Feynman rules\label{Feynman rules}}

Let us describe the rules for diagrams. We use the definition of metrics and Dirac matrices by S.~Weinberg \cite{W3}:
\begin{equation}
\eta_{\mu\nu}=\mbox{diag}(-1,+1,+1,+1),\qquad
\{\gamma_\mu,\gamma_\nu\}=2\eta_{\mu\nu}.
\end{equation}

The propagator of inflaton is denoted by the solid line corresponding to the expression
\begin{center}
\includegraphics{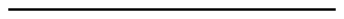}
\end{center}

\begin{equation}
(-i)\frac{1}{p^2+m^2},
\end{equation}
where $m$ is the inflaton mass.

The inflatino is denoted by the double line with the propagator of the form
\begin{center}
\includegraphics {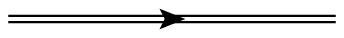}
\end{center}

\begin{equation}
(-i)\frac{-i\gamma p+m'}{p^2+m'^2},
\end{equation}
where $m'$ stands for the inflatino mass.

The gravitino is shown by the dashed line with the propagator
\begin{center}
\includegraphics {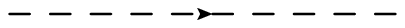}
\end{center}

\begin{eqnarray}
(-i)\frac{P^{\mu\nu}(p)}{p^2+m_g^2} &=&
(-i)\frac{1}{p^2+m^2_g}\left\{
\left(\eta^{\mu\nu}+\frac{p^\mu p^\nu}{m^2_g}\right)(-i \gamma p+m_g)
\right.\\
\nonumber &&
\left.
-\frac13
\left(\gamma^\mu-i\frac{p^\mu}{m_g}\right)(i\gamma p+m_g)\left(\gamma^\nu-i
\frac{p^\nu}{m_g}\right)
\right\},
\end{eqnarray}
where $m_g=\sqrt{\frac{(2\pi G}{3} \Lambda^4}$ is the gravitino mass.

The vertexes with the gravitino appear by involving the supersymmetry so that the gravitino interacts with the supercurrent through the term of lagrangian
\begin{equation}
\sqrt{8\pi G}\int \ud ^4 x \frac12 \overline{S}^\mu \psi _\mu,
\end{equation}
where $\psi^\mu$ denotes the gravitino field, while the supercurrent equals
$$S^\mu=\sqrt2 \left[
\gamma^\nu \partial_\nu \phi \gamma^\mu \psi _R + \gamma^\nu
\partial_\nu \phi ^* \gamma^\mu \psi _L + \left(\frac{\partial W}{\partial
\phi}\right)\gamma^\mu \psi_L
+\left(\frac{\partial W}{\partial \phi}\right)^* \gamma^\mu \psi_R
\right],$$ here $\phi$ denotes the inflaton, $\psi$ does the inflatino, and $W$ is the superpotential.

The lagrangian contains the potential terms corresponding to the inflaton self-action and the coupling of inflaton to the inflatino
\begin{equation}
\int \ud x^4\left\{ \frac12 \left(\frac{\partial^2 W}{\partial \Phi
^2}\right)(\overline{\psi_L}\psi_L)+\frac12 \left(\frac{\partial^2
W}{\partial \Phi ^2}\right)^*(\overline{\psi_L}\psi_L)^*+\left(
\frac{\partial W}{\partial\Phi}\right)\left( \frac{\partial
W}{\partial\Phi}\right)^*\right\}.
\end{equation}

There are three kinds of vertexes appearing from the supercurrent:
\begin{itemize}
\item three lines of inflaton, inflatino and gravitino come into the first vertex, so that we get

\begin{center}
\includegraphics {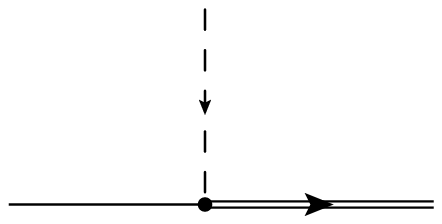}
\end{center}
\begin{equation}
(i) (i) \frac12 \mu_0\sqrt{8\pi G} \gamma^\nu \gamma^5 =
-\frac12\mu_0\sqrt{8\pi G} \gamma^\nu \gamma^5,
\end{equation}
where $\gamma^5=i\gamma^0\gamma^1\gamma^2\gamma^3$;

\item four lines of inflatino, gravitino and twice inflaton due to its self-action compose the second vertex
\begin{center}
\includegraphics {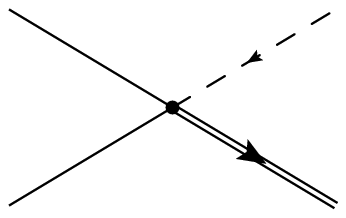}
\end{center}
\begin{equation}
(i) \frac1{\sqrt2} g_0\sqrt{8\pi G} \gamma^\nu,
\end{equation}

\item the inflatino is transformed into the gravitino in the third vertex

\begin{center}
\includegraphics {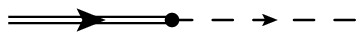}
\end{center}

\begin{equation}
i\frac1{\sqrt2}\sqrt{8\pi G}\Lambda^2\gamma^\nu.
\end{equation}

\end{itemize}

The propagator of graviton looks like
\begin{center}
\includegraphics {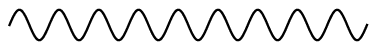}
\end{center}

\begin{equation}
G^{\mu\nu , \mu'\nu'}=(-i)\frac{1}{2p^2}(
\eta^{\mu\mu'}\eta^{\nu\nu'}+\eta^{\mu\nu'}\eta^{\nu\mu'}-\eta^{\mu\nu}\eta^{
\mu\nu'}).
\end{equation}

The vertex of coupling the graviton to the scalar particle is determined by the term of lagrangian $\sqrt{8\pi G}\int \ud ^4 x T^{\mu\nu} h_{\mu\nu}$, where
$T^{\mu\nu}$ is the tensor of energy-momentum, while $h_{\mu\nu}$ is given by $g_{\mu\nu}=\eta_{\mu\nu}+2\sqrt{8\pi G}h_{\mu\nu}$, so that the vertex is equal to

\begin{center}
\includegraphics {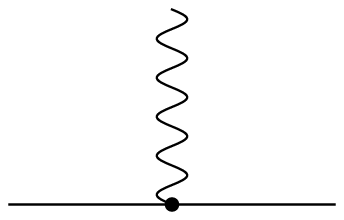}
\end{center}

\begin{equation}
(-i) \sqrt{8\pi G} \eta^{\mu\nu} (\mu_0^2+2g_0\Lambda^2).
\end{equation}

The quartic self-action gives the vertex
\begin{center}
\includegraphics {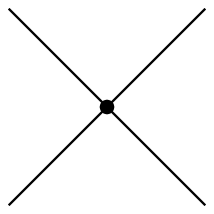}
\end{center}

\begin{equation}
-i \lambda_0 = -6i g_0^2,
\end{equation}
and the vertex for the interaction between the inflaton and inflatino appears from the lagrangian after the chiral rotation (Appendix \ref{kirrot}):

\begin{center}
\includegraphics {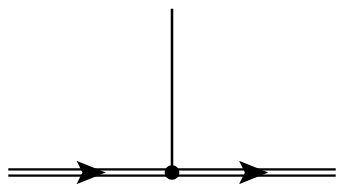}
\end{center}

\begin{equation}
- \sqrt2 g_0 \gamma^5.
\end{equation}

All of vertexes preserve the conservation of momentum. The indefinite momenta are integrated out as $\frac{\ud^4 p}{(2\pi)^4}$. Each fermionic loop makes the factor $(-1)$. The Wick rotation corresponds to the substitution of  $p_0 \to i
p_4 $ in the transformation to the Euclidean coordinates.

\end{document}